\documentclass[12pt]{article}
\usepackage{amsmath} \usepackage{amsfonts} \usepackage{amssymb}
\usepackage{mathtools}
\usepackage{bbm}
\usepackage{epsfig}
\usepackage{graphicx}
\usepackage{float}
\usepackage[shortlabels]{enumitem}
\usepackage{mathabx}
\usepackage{caption}
\usepackage{subcaption}
\usepackage{subfig}
\usepackage{verbatim}
\usepackage{multirow}
\usepackage{cite}
\usepackage{lscape}
\usepackage{feynmp}
\usepackage[title]{appendix}
\usepackage{tikz}
\usetikzlibrary{matrix,positioning}
\newcommand{\be}{\begin{equation}}
\newcommand{\ee}{\end{equation}}
\newcommand{\ba}{\begin{eqnarray}}
\newcommand{\ea}{\end{eqnarray}}
\newcommand{\nn}{\nonumber}
\newcommand{\mbf}[1]{\mathbf{#1}}
\newcommand{\mbv}[2]{\mathbf{#1}_{#2}}
\newcommand{\gam}{\gamma^*}
\newcommand{\rhog}[1]{\hat\rho^{(G,G)}_{#1}}
\newcommand{\xx}{(x_1,x_2)}
\newcommand{\braket}[3]{\langle #1 \left|#2 \right| #3 \rangle}

\numberwithin{equation}{section} 
\begin{document}

\title{\bf Spin correlations in the Drell-Yan process, parton entanglement, and other 
unconventional QCD effects}

\author{\bf O. Nachtmann$^1$\\[3mm]
Institut f\"ur Theoretische Physik\\
Universit\"at Heidelberg\\
Philosophenweg 16,
D-69120 Heidelberg, Germany}
\date{}
\maketitle
\thispagestyle{empty}

\begin{abstract}
We review ideas on the structure of the QCD vacuum which had served as motivation for the 
discussion of various non-standard QCD effects in high-energy reactions in articles from 1984 to 1995.
These effects include, in particular, transverse-momentum and spin correlations in the Drell-Yan 
process and soft photon production in hadron-hadron collisions. 
We discuss the relation of the approach introduced in the above-mentioned articles to the approach, 
developed later, using transverse-momentum-dependent parton distributions (TDMs).
The latter approach is a special case of our more general one which allows for parton entanglement in 
high-energy reactions. We discuss signatures of parton entanglement in the Drell-Yan reaction.
Also for Higgs-boson production in $pp$ collisions via gluon-gluon annihilation effects of entanglement of 
the two gluons are discussed and are found to be potentially important.
These effects can be looked for in the current LHC experiments. In our opinion studying 
parton-entanglement effects in high-energy reactions is, on the one hand, very worthwhile by 
itself and, on the other hand, 
it allows to perform quantitative tests of standard factorisation assumptions. Clearly, the experimental 
observation of parton-entanglement effects in the Drell-Yan reaction and/or in Higgs-boson production would have 
a great impact on our understanding how QCD works in high-energy collisions.
\end{abstract}

\vspace{13em}
\hrule width 5.cm
\vspace*{.5em}
{\small \noindent
$^1$ email: O.Nachtmann@thphys.uni-heidelberg.de}
\newpage

\setcounter{page}{1}

\section{Introduction}
\label{Introduction}
In this article we want to give a synopsis and an update of the results of \cite{1,2,3,4,5} concerning 
some unconventional QCD effects in the Drell-Yan process and in soft-photon production in hadron-hadron 
collisions. 
In addition we shall investigate possible effects of parton entanglement for Higgs-boson production in 
hadron-hadron collisions. 
We think that our study is quite timely. On the one hand there are the current LHC experiments. 
On the other hand there is an experimental program under way to investigate over a 
large c.m.\ energy range the Drell-Yan process and the related $Z$-production reaction
\begin{align}\label{1.1}
 \begin{split}
h_1+h_2 \to & V+X,\\
  & \drsh l^+ + l^-\\
V  = &\,\gamma^*,Z.\\
\end{split}
\end{align}
Here $h_{1,2}$ are hadrons, $X$ stands for the final hadronic state, and $l=e,\mu$ for the leptons. 
We shall be interested in particular in the angular distribution of the leptons where ``anomalies'' 
have first been seen by the NA10 experiment at CERN \cite{6,7} and then confirmed by the E615 
experiment at FNAL \cite{8,9}. 
The interesting findings of these experiments have only recently led to great further experimental 
efforts. In table 1 we list the original experiments and some recent ones which are either completed 
or planned. This list is {\em not} intended to be exhaustive, it is only meant to indicate the 
wide range of ongoing studies, concerning both the incoming hadrons $h_{1,2}$ in \eqref{1.1} and the 
c.m.\ energy $\sqrt{s}$. 
All these experiments should be very suitable for studying the unconventional QCD effects 
discussed in \cite{1,2,3,4,5}.  

Our paper is organised as follows. In section \ref{The QCD vaccum structure} we recall some 
ideas on the QCD vacuum structure which were developed in the 1970s and 1980s. 
We sketch the motivation which led to the introduction of spin correlations in the Drell-Yan 
process in \cite{1,2}. In section \ref{Transverse Momentum} 
we discuss the framework developed in \cite{2} for treating the reaction \eqref{1.1}. 
The relation of our framework to the one using transverse-momentum-dependent-parton 
distributions (TMDs) is given. 
We emphasise that our framework allows to investigate effects from parton entanglement 
which may occur, for instance, due to instantons. 
In section \ref{Higgs-boson production} we investigate possible effects of parton 
entanglement - in this case for gluons - on the production of Higgs bosons in hadron-hadron collisions. 
Section \ref{Conclusions} contains our conclusions. 
In appendices we discuss the Drell-Yan reaction with general quark-antiquark density matrix, 
conventions for kinematic variables, and an example of a non-trivial two-gluon density matrix for Higgs-boson 
production via gluon-gluon annihilation for entangled gluons. 

\begin{center}
\begin{table}[h!] 
\caption{The parameters of the experiments NA10 and E615 from the late 1980s and a partial 
list of recently completed and planned experiments for reaction \eqref{1.1}. The approximate c.m.\ 
energies $\sqrt{s}$ and years of running are also indicated. This table is in part based 
on material presented in \cite{21}.} 
\vspace{0.5cm}
\begin{tabular}{c|c|c|c|c|c}
Experiment&$h_1$&$h_2$&$V$&$\sqrt{s}$ [GeV]&years\\ \hline \hline
NA10 (CERN)&$\pi^-$&$W,d$&$\gamma^*$&$16$ to $23$&1986-1988\\
$[6,7]$&&&&\\
&&&&\\
E615 (FNAL)&$\pi^-$&$W$&$\gamma^*$&$22$&1989-1991\\
$[8,9]$&&&&\\ \hline \hline
PANDA (GSI)&$\bar p$&$p$&$\gamma^*$&$5.5$&$>2016$\\
$[10]$&&&&\\
&&&&\\
PAX (GSI)&$\bar p$&$p$&$\gamma^*$&$14$&$>2017$\\
$[11]$&&&&\\
&&&&\\
E906 (FNAL)&$p$&$p$&$\gamma^*$&$15$&2011\\
$[12]$&&&&\\
&&&&\\
COMPASS II (CERN)&$\pi^\pm$&$p$&$\gamma^*$&$17.4$&2014\\
$[13]$&&&&\\
&&&&\\
E866 (FNAL)&$p$&$p,d$&$\gamma^*$&$39$&2007-2009\\
$[14,15]$&&&&\\
&&&&\\
PHENIX (BNL)&$p$&$p$&$\gamma^*$&$200$&$>2018$\\
$[16]$&&&&\\
&&&&\\
CDF (FNAL)&$\bar p$&$p$&$\gamma^*,Z$&$1.96\times 10^3$&2011\\ 
$[17]$&&&&\\
&&&&\\
$\left.\begin{array}{l}
 \text{ALICE}\\\text{ATLAS}\\\text{CMS}
\end{array}\right\} \text{(CERN)}$&$p$&$p$&$\gamma^*,Z$&$2\times 10^3$ to $14\times 10^3$&2010-\\
$[18-20]$&&&&\\
\end{tabular}
\end{table}
\end{center}

\section{The QCD vacuum structure as a possible source of unconventional effects}
\label{The QCD vaccum structure}
In the 1970s and 1980s many interesting ideas on the QCD vacuum structure were developed. 
Instantons were introduced and shown to have important effects in \cite{110,111,112}. 
Savvidy \cite{113} showed that a colour-magnetic field will lower the energy of the vacuum state. 
Shifman, Vainshtein and Zakharov (SVZ) introduced the gluon condensate of the vacuum \cite{114,115,115a}. 
A particularly nice picture of the QCD vacuum was developed by Ambj{\o}rn and Olesen \cite{116a,116}; see 
figure 1a. Note the hexagonal structure of the chromomagnetic flux tubes. For comparison we show in 
figure 1b the hexagonal structure of the ether of electrodynamics as envisaged by Maxwell \cite{117} 
in 1861. Thus, some ideas on the QCD vacuum structure resemble strikingly the dielectric ether of the 
19th century which was discarded by Einstein. Maybe, some time in the future we shall also have a 
deeper and simpler understanding of the QCD vacuum structure. For the present we shall take these 
ideas on the QCD vacuum as working hypothesis; 
for reviews see \cite{118} and chapters 6 and 8 of \cite{119}. 

\begin{figure}
\begin{subfigure}[H]{.5\linewidth}
  \centering
  \includegraphics[width=1.0\textwidth]{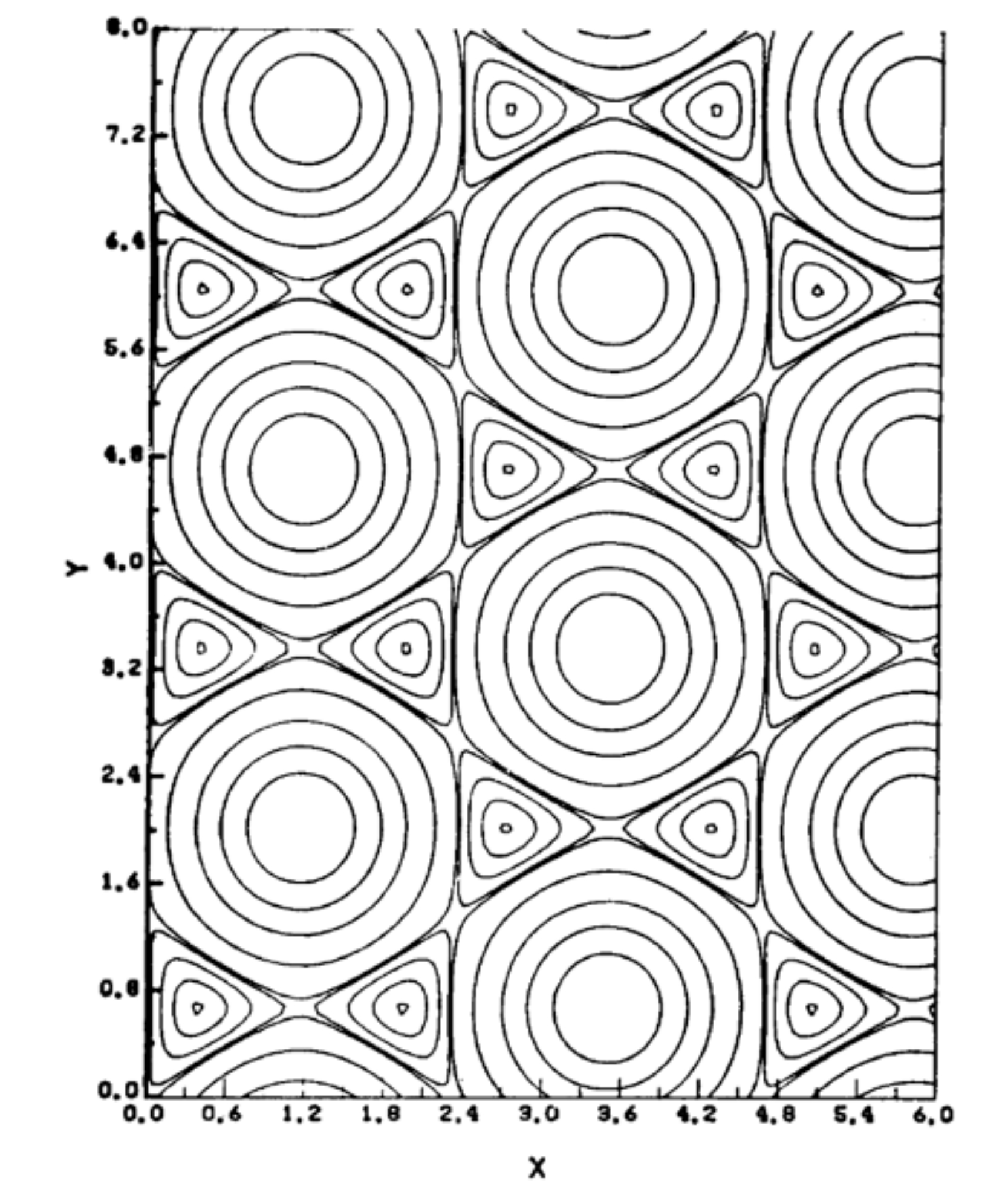}\\(a)
  \label{SCD_fig11}
\end{subfigure}
\begin{subfigure}[H]{.5\linewidth}
  \centering
  \includegraphics[width=1.0\textwidth]{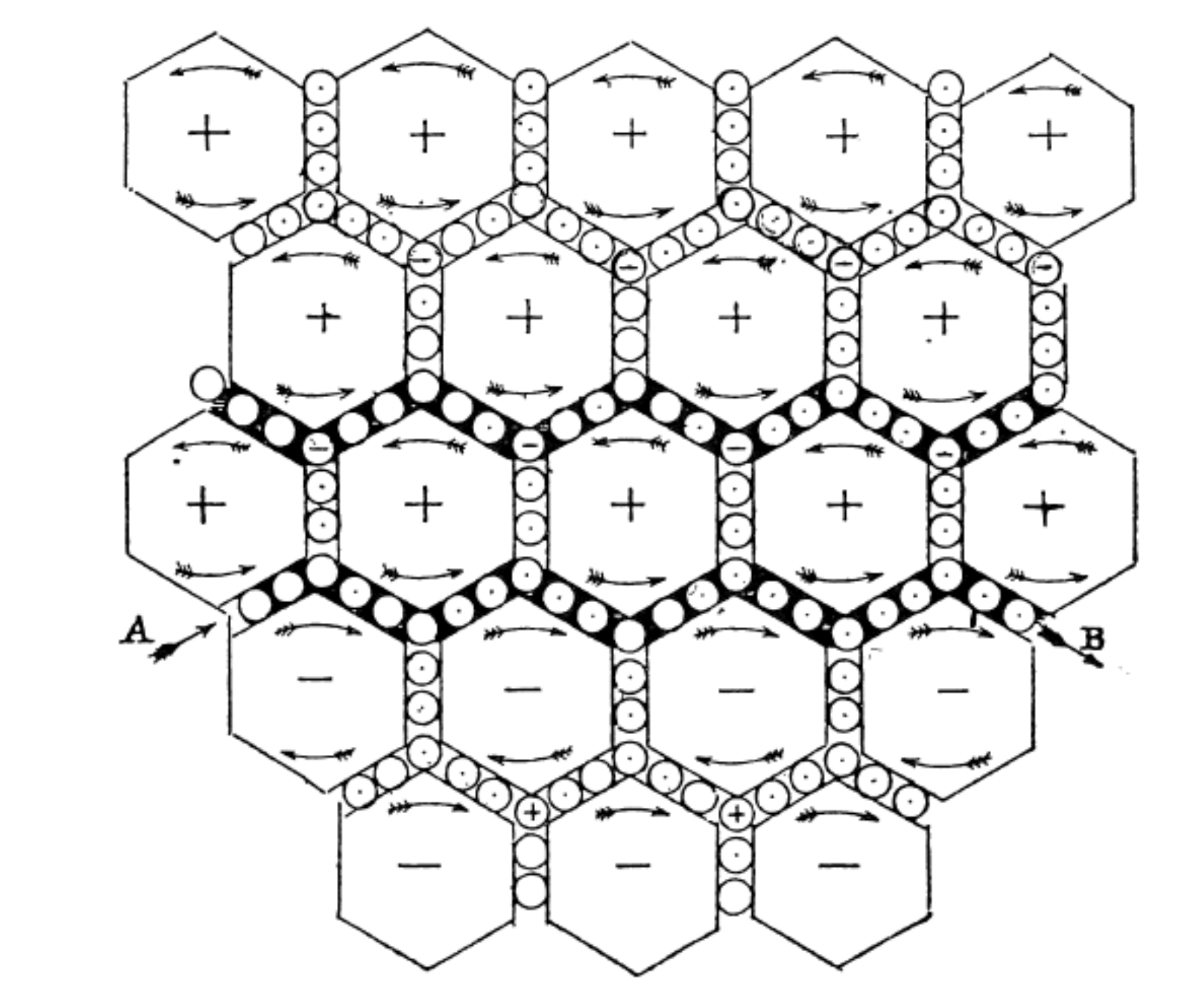}\\(b)
  \label{fig:one-two}
\end{subfigure}
\caption{The QCD vacuum according to Ambj{\o}rn and Olesen \cite{116} (a); 
the ether according to Maxwell \cite{117} (b).}
\label{SCD_fig12}
\end{figure}

But let us come to the Drell-Yan (DY) reaction \eqref{1.1} with $V=\gamma^*$. In leading order we 
have the annihilation of a quark-antiquark pair into a virtual photon $\gamma^*$ which then decays 
into a lepton pair; see figure 2. The standard description of the process is well 
known \cite{120,121,121a}. For unpolarised hadrons $h_1$ and $h_2$ the quark $q$ and 
antiquark $\bar q$ are supposed to be also unpolarised and completely uncorrelated. 

\begin{figure}[htb]
\begin{center}
\includegraphics[width=0.65\textwidth]{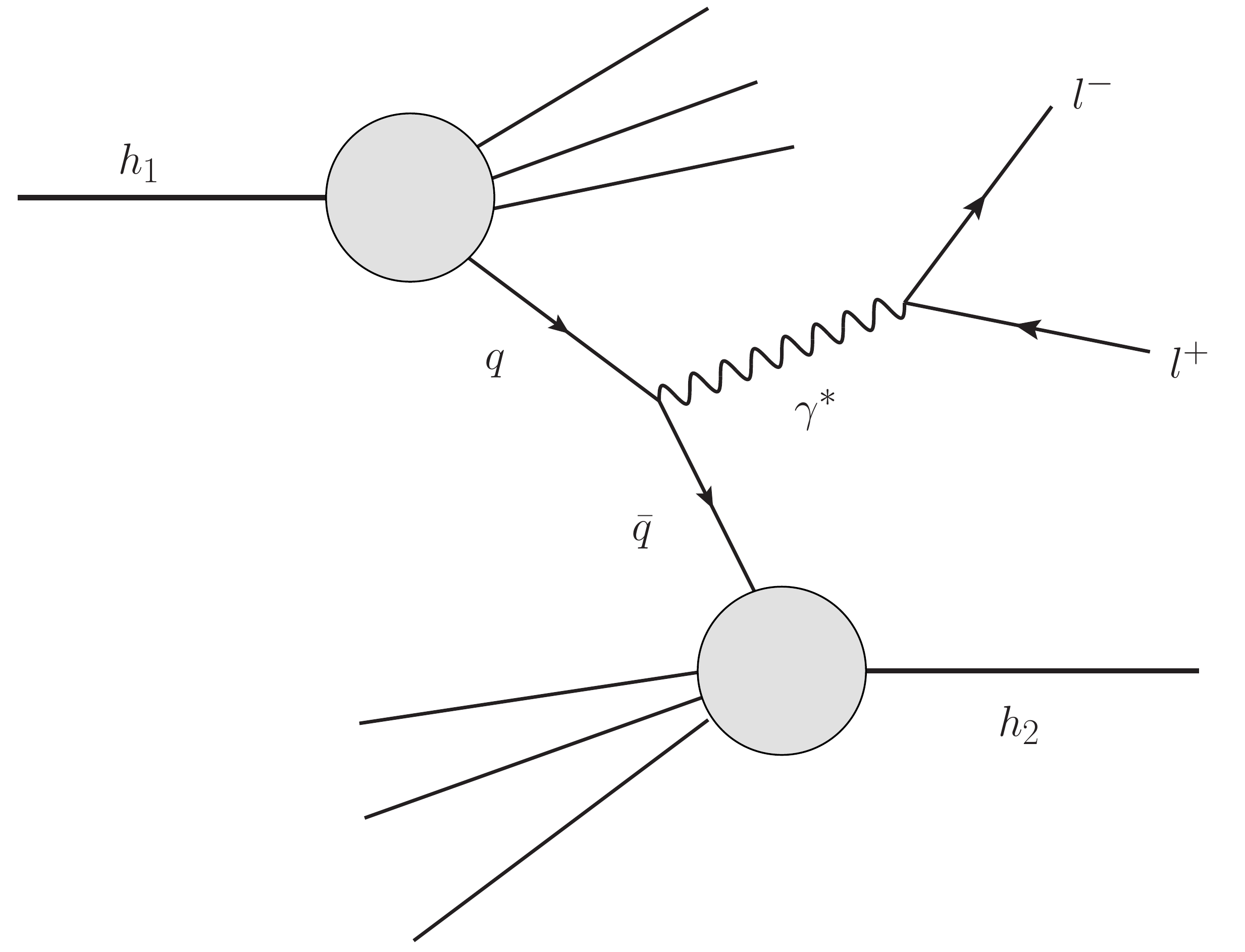}
\caption{Leading order diagram for the Drell-Yan process \eqref{1.1} with $V=\gamma^*$, $q$ from $h_1$ and $\bar q$ from $h_2$. The diagram with the role of $q$ and $\bar q$ exchanged has to be added.}
\label{SCD_fig2}
\end{center}
\end{figure}

In the paper\cite{1} of 1984 this assumption was questioned. The argument was that the $q\bar q$ 
annihilation takes place in a non-trivial background full of colour fields if we believe the ideas 
on the QCD vacuum. It was argued that this could give rise to spin and even colour-spin correlations 
of $q$ and $\bar q$, and possibly to {\em entanglement} of the two partons. Before we recall these 
arguments we want to mention that already in \cite{122} it was shown that instanton effects in the 
DY process will {\em not} go away at high energies. But these authors only considered the total rates 
and were not concerned with spin effects. In fact, they only considered spinless partons. 

Let us start with the gluon condensate of SVZ \cite{114,115,115a}. From Lorentz and parity invariance of 
the strong interactions we find for the vacuum expectation value of the product of two gluon field 
strengths at the same space-time point 
\ba\label{2.1}
\langle 0|\frac{g^2_s}{4\pi^2}:G^a_{\mu\nu}(x)
G^b_{\rho\sigma}(x):|0\rangle 
=\frac{1}{96}\delta^{ab}(g_{\mu\rho}g_{\nu\sigma}-g_{\mu\sigma}g_{\nu\rho})G_2.
\ea
Here $g_s$ is the QCD coupling constant and $a,b,\in \{1,\dots ,8\}$ are the colour indices. 
A typical phenomenological value for $G_2$ is 
\be\label{2.2}
G_2=\frac{1}{4\pi^2}(0.95\pm 0.45) \text{ GeV}^4;
\ee
see \cite{123}. For further discussion of the value of $G_2$ see for instance \cite{124,125}. 
For the chromoelectric and chromomagnetic fields we get from \eqref{2.1} 
\ba\label{2.3}
-\langle 0|g^2_s:\mathbf E^a(x)~\mathbf E^a(x):|0\rangle
&=&\langle 0|g^2_s:\mathbf B^a(x)~\mathbf B^a(x):|0\rangle\nn\\
&=&\pi^2 G_2 ~ \widetilde{ = } ~ (700 \text{ MeV})^4.
\ea
One assumes that the products of the fields in \eqref{2.1} and \eqref{2.3} are normal ordered 
with respect to the ``perturbative vacuum state'', whatever this is. The interpretation of \eqref{2.3} 
should, therefore, be that in the physical vacuum the chromomagnetic field fluctuates with larger 
amplitude, the chromoelectric field with smaller amplitude than in the perturbative vacuum. 
In \cite{1,2,3,4} possible effects from these strong chromomagnetic vacuum fields were discussed. 

\begin{figure}[htb]
\begin{center}
\includegraphics[width=0.65\textwidth]{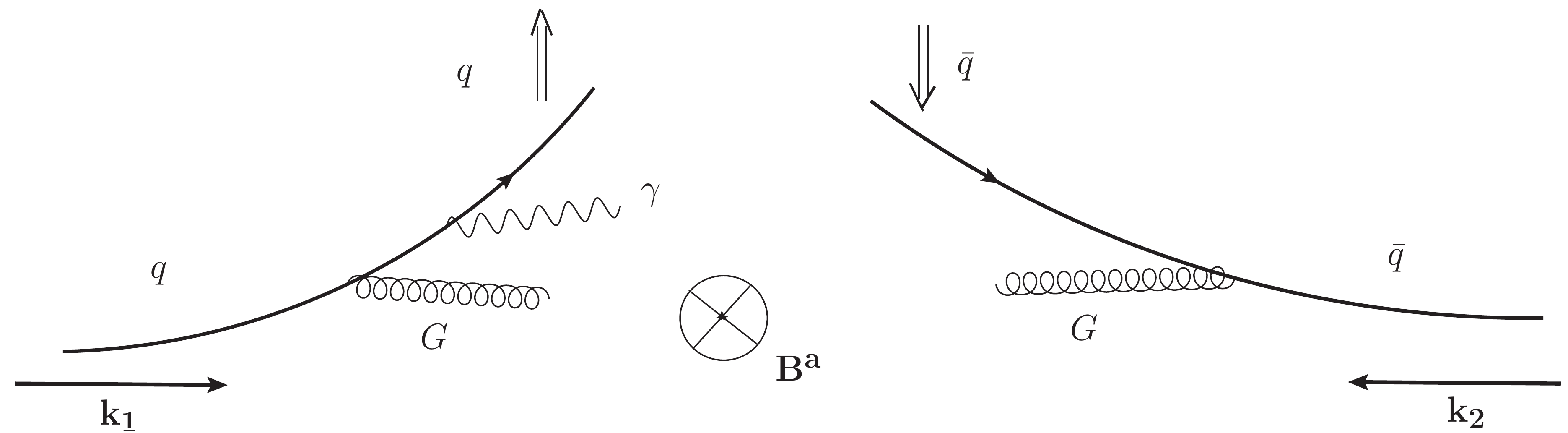}
\caption{A quark and an antiquark with momenta $\mathbf{k_1}$ and $\mathbf{k_2}$, respectively, moving through 
a chromomagnetic background field.}
\label{SCD_fig3}
\end{center}
\end{figure}

Let us envisage a light quark $u$ or $d$ with mass of a few MeV moving in a constant chromomagnetic 
background field of the strength given in \eqref{2.3}. What will happen? The quark will be deflected 
due to the chromomagnetic Lorentz force and perform a cyclotron motion much like an electron in a 
storage ring; see figure \ref{SCD_fig3}. The quark will emit synchrotron gluons, and since it also 
carries electric charge, also synchrotron photons. The quark will also emit spin-flip synchrotron 
gluons and by this get a transverse polarisation. This is the analogon of the well known 
Sokolov-Ternov effect in storage rings \cite{126,127,128}. 

If now in addition an antiquark sails through the same background field it also will get deflected, 
emit synchrotron gluons and photons and will get a polarisation. If all this happens 
for $q$ and $\bar q$ in the {\em same background field} they will develop a correlation, both, 
in transverse momenta and spins. The background field may thus be a source of {\em parton entanglement}. 

\begin{figure}[htb]
\begin{center}
\includegraphics[width=0.65\textwidth]{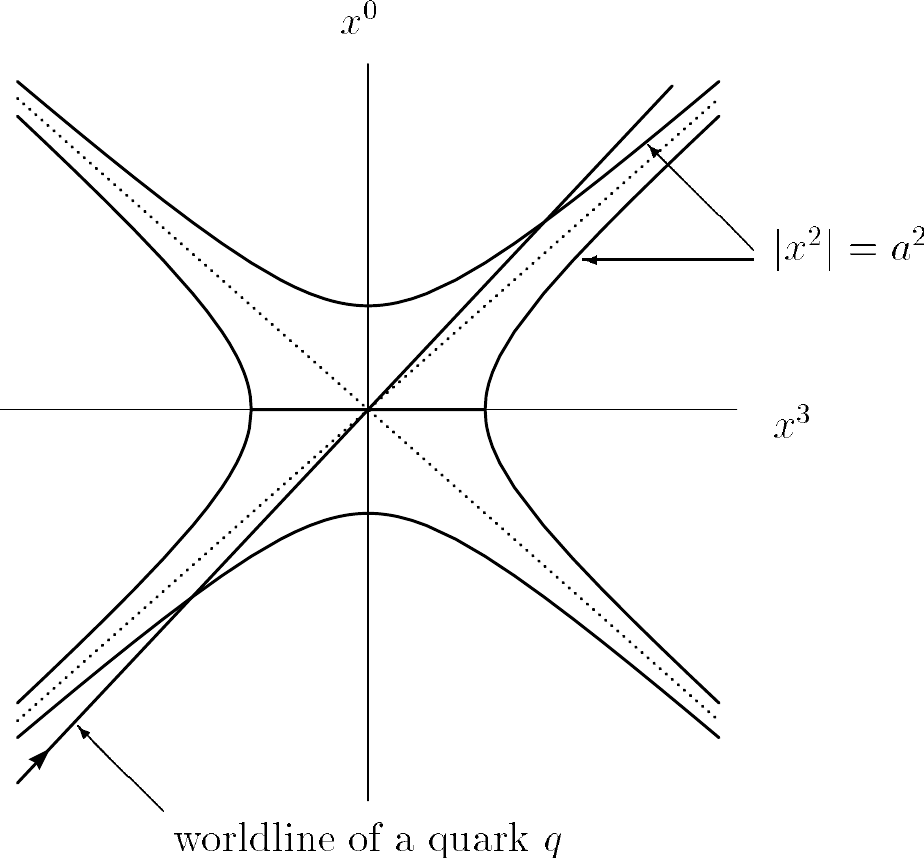}
\caption{Sketch of a correlation region in Minkowski space with the world line of a fast quark 
passing through it (from figure 1 of \cite{3}).}
\label{SCD_fi4}
\end{center}
\end{figure}

Of course, there can be no constant background chromomagnetic field in the vacuum. But there can 
be {\em correlations} of fields at different space-time points. Indeed, in Euclidean QCD one finds 
from lattice calculations, see  for instance \cite{129,125}, that such correlations fall off 
exponentially with a typical correlation length 
\be\label{2.4}
a ~ \widetilde{=}~ 0.2 \text{ to } 0.3 \text{ fm}.
\ee
That is, the gluon field strengths are highly correlated for points of Euclidean separation 
$x_{\text{Eucl.}}$ satisfying
\be\label{2.5}
x^2_{\text{Eucl.}}\lesssim a^2.
\ee
An instanton enthusiast may think of $a$ as the typical size of instantons contributing to our 
effects. For the size distribution of instantons see for instance \cite{42a,42b}.
Translating this in the most straightforward way to Minkowski space we find that \eqref{2.5} 
implies a strong correlation of fields separated by a Minkowskian distance $x_{\text{Mink.}}$ with 
\be\label{2.6}
|x^2_{\text{Mink.}}|\lesssim a^2.
\ee
A sketch of such a region, centred at $x_{\text{Mink.}}$ = 0 is shown in figure \ref{SCD_fi4}. 

\begin{figure}[htb]
\begin{center}
\includegraphics[width=0.65\textwidth]{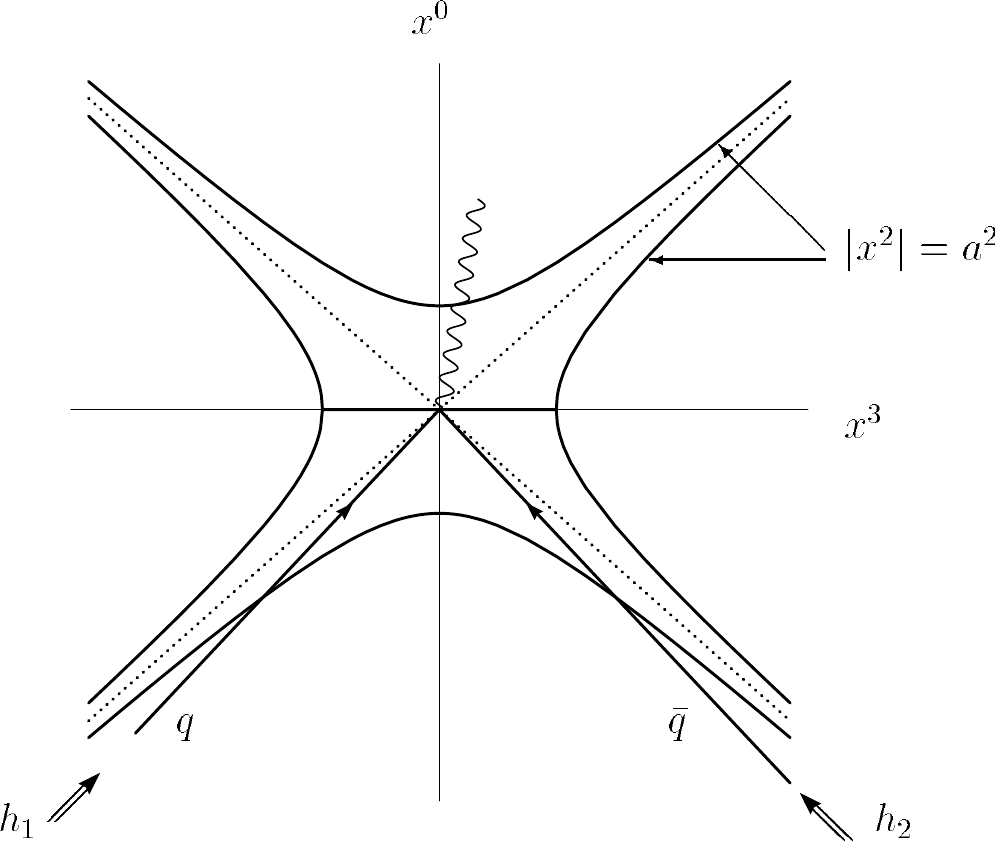}
\caption{Annihilation of a quark $q$ and antiquark $\bar q$ from two hadrons $h_1$ and $h_2$, 
respectively, with production of a virtual photon $\gamma^*$. Before the annihilation hard 
partons $q$ and $\bar q$ spend a long time in a correlated field region (from figure 2 of \cite{3}).}
\label{SCD_fig5}
\end{center}
\end{figure}

What can we say from these considerations for the Drell-Yan process shown in figure \ref{SCD_fig2}? 
Suppose that the fast quark $q$ and antiquark $\bar q$ from hadrons $h_1$ and $h_2$, respectively, 
annihilate at point $x=0$ in Minkowski space; see figure \ref{SCD_fig5}. 
The $q$ and $\bar q$ will then have the chance to spend a {\em long time} in a correlated domain. 
Explicit estimates show that, indeed, there is enough time for transverse-momentum, 
spin-transverse-momentum, 
and spin-spin correlations of hard partons $q$ and $\bar q$ to develop; see \cite{1,2,3}. 

With these remarks we shall end our short review of the ideas on the QCD vacuum structure and how 
this gave motivations in \cite{1,2,3,4} to discuss various unconventional QCD effects for 
high-energy processes. 
The picture emerging from these considerations can be summarised in the following points. 
\newpage

\bigskip\noindent
{\bf Summary of section 2:}
\begin{itemize}
\item [(i)] The quarks, and similarly hard gluons, of a fast hadron $h$ feel a background 
chromomagnetic field of typical strength $g_sB_c$. The chromomagnetic Lorentz force causes 
the coloured partons to ``wiggle'' and to emit ordinary and spin-flip synchrotron gluons. 
Quarks, being charged, also emit synchrotron photons. Of course, for an isolated hadron $h$ there 
can be no real radiation. 
The synchrotron gluons and photons are part of the cloud of virtual particles around the hadron.
These soft effects should affect the transverse-momentum and the spin distributions of the hard partons 
but should not influence their longitudinal-momentum distributions appreciably.
\item[(ii)] In \cite{3} it was estimated that different hard partons of the hadron $h$ travel 
generally in uncorrelated field domains. This implies, for instance, that we should add the 
synchrotron photons emitted from various hard partons in the hadron incoherently. 
\item[(iii)] In a hadron-hadron collision the cloud of synchrotron photons may be shaken off 
and should give rise to soft photons with a characteristic energy spectrum, see \cite{1,3}. 
The main parameter governing this soft-photon yield turned out to be an effective  length $l_{\text{eff}}$  
defined as the distance a fast quark has to travel in the chromomagnetic field of strength $g_sB_c$ 
for obtaining the typical transverse momentum $\bar p_T$ which quarks have in the hadron. 
The ordinary cyclotron formulae lead to the estimate
\be\label{2.7}
l_{\text{eff}}~\widetilde{=}~\frac{\bar p_T}{g_sB_c}.
\ee
From a comparison of the theory with the experimental results on soft photons in $p-Be$ 
collisions of \cite{130} a value of 
\be\label{2.8}
l_{\text{eff}}~\widetilde{=}~20 \text{ to }40 \text{ fm}
\ee
was extracted in \cite{3}. With $\bar p_T~\widetilde{=}~300$ MeV, a typical value for the mean 
transverse momentum of quarks in hadrons, we obtain from \eqref{2.7} for the effective chromomagnetic 
background field in a hadron
\ba\label{2.9}
g_sB_c~\widetilde{=}~ (44 \text{ MeV})^2,
\text{ for }l_{\text{eff}}=30 \text{ fm}.
\ea
This is much smaller than the strength of the vacuum fields in \eqref{2.3}: 
\be\label{2.10}
g_sB_{\text{vac}}~\widetilde{=}~(700\text{ MeV})^2.
\ee
A possible resolution of this puzzle was suggested in \cite{3,4}: the vacuum fields must be shielded 
by gluons, maybe those from the synchrotron effects, otherwise quarks in a fast hadron could not get 
far without being strongly bent. Indeed, inserting $g_sB_{\text{vac}}$ instead of $g_sB_c$ 
in \eqref{2.7} we find $l_{\text{eff}}~\widetilde{=}~0.1$ fm which, in our opinion, is ridiculously 
small. Thus, in this view gluons in a fast hadron play an important dynamical role: they have to 
shield the vacuum chromomagnetic fields in order to allow quarks to travel more or less on straight 
paths. 
\item[(iv)] The gluonic spin-flip-synchrotron radiation of quarks in the chromomagnetic background 
field may also have some relevance for the proton spin puzzle; see \cite{131} for a recent review. 
Indeed, consider a fast proton with helicity $+1/2$. This angular momentum must be built up by the 
spin and orbital angular momenta of the constituents, the partons, 
\ba\label{2.11}
\frac12 \Delta\Sigma +L_q+\Delta G+L_g=\frac12.
\ea
Here $\Delta\Sigma/2$ and $L_q$ are the contributions from the spin and orbital angular 
momenta, respectively, of quarks and antiquarks. The corresponding contributions from the gluons 
are denoted by $\Delta G$ and $L_g$. It is known for about 25 years that 
\be\label{2.12}
\Delta\Sigma ~\widetilde{=}~0.25\ll 1;
\ee
see \cite{131}. The puzzle posed by \eqref{2.12} is why the spin of the quarks contributes so little to the 
proton helicity.  In our framework we can argue that in the chromomagnetic background field an original 
longitudinal polarisation of a quark in the fast proton will be degraded and be partly turned into 
a transverse one by gluon spin-flip synchrotron radiation. This is in complete analogy to what happens 
in $e^+e^-$ storage rings due to the Sokolov-Ternov effect \cite{126,127,128}. The expectation is 
then that rather soft gluons will guarantee the angular momentum balance in these spin-flip processes 
and thus in \eqref{2.11}. In experiments on the proton tomography being currently 
discussed \cite{44b,44c} one may be able to check these ideas, originally put forward in \cite{3,4} 
and summarised here as points (iii) and (iv). 

\item[(v)] For the Drell-Yan process, see \eqref{1.1} and figure \ref{SCD_fig2}, transverse-momentum, 
spin-transverse-momentum, and spin-spin correlations of the annihilating quark-antiquark pair were 
predicted in \cite{1,2} and were shown to lead to ``anomalous'' effects in the angular distribution 
of the produced leptons. 
Such anomalies were indeed observed by experiments; see \cite{6,7,8}. We shall discuss all this 
in some detail in section \ref{Transverse Momentum} below. 
In section \ref{Higgs-boson production} we will discuss gluon-gluon spin correlations and their 
effects on Higgs-boson production via gluon-gluon fusion. 
\end{itemize}

\section{Transverse-momentum, spin-transverse-momentum,\\ and spin-spin correlations 
in the Drell-Yan process}
\label{Transverse Momentum}
In this section we consider the general Drell-Yan process, that is, the production of a virtual 
photon $\gamma^*$ or $Z$ boson in a hadron-hadron collision 
\be\label{3.1}
h_1(p_1)+h_2(p_2)\to V(k,\epsilon)+X~,~V=\gamma^*,Z.
\ee
Completely analogous is $W^\pm$ production 
\be\label{3.2}
h_1(p_1)+h_2(p_2)\to W^\pm(k,\epsilon)+X.
\ee
Here we shall consider explicitly only $\gamma^*$ and $Z$ production. For a discussion of $W^\pm$ 
production from our point of view we refer to \cite{2}.
In \eqref{3.1} and \eqref{3.2} the momenta are indicated in brackets and $\epsilon$ is the 
polarisation vector of the vector boson. We will only discuss the case of unpolarised hadrons 
$h_1,h_2$ and the leading order process where a quark-antiquark pair annihilates to give $V$ 
in \eqref{3.1} 
\be\label{3.3}
q(k_1)+\bar q(k_2)\to V(k,\epsilon).
\ee
For massless quarks their momenta in the overall c.m.\ system are
\be\label{3.4}
k_1=\left(\begin{array}{c}
\sqrt{x^2_1{\mathbf p}^2_1+{\mathbf k}^2_{1T}}\\
x_1{\mathbf p}_1+{\mathbf k}_{1T}
\end{array}\right)
~,~
k_2=\left(\begin{array}{c}
\sqrt{x^2_2{\mathbf p}^2_2+{\mathbf k}^2_{2T}}\\
x_2{\mathbf p}_2+{\mathbf k}_{2T}
\end{array}\right)
\ee
where $x_1(x_2)$ is the fraction of longitudinal momentum of $h_1(h_2)$ carried by 
quark $q$ (antiquark ${\bar q}$). Of course, in the following it is always understood 
that one also has to consider the case where ${\bar q}$ comes from $h_1$ and $q$ from $h_2$. 

\begin{figure}[htb]
\begin{center}
\includegraphics[width=0.65\textwidth]{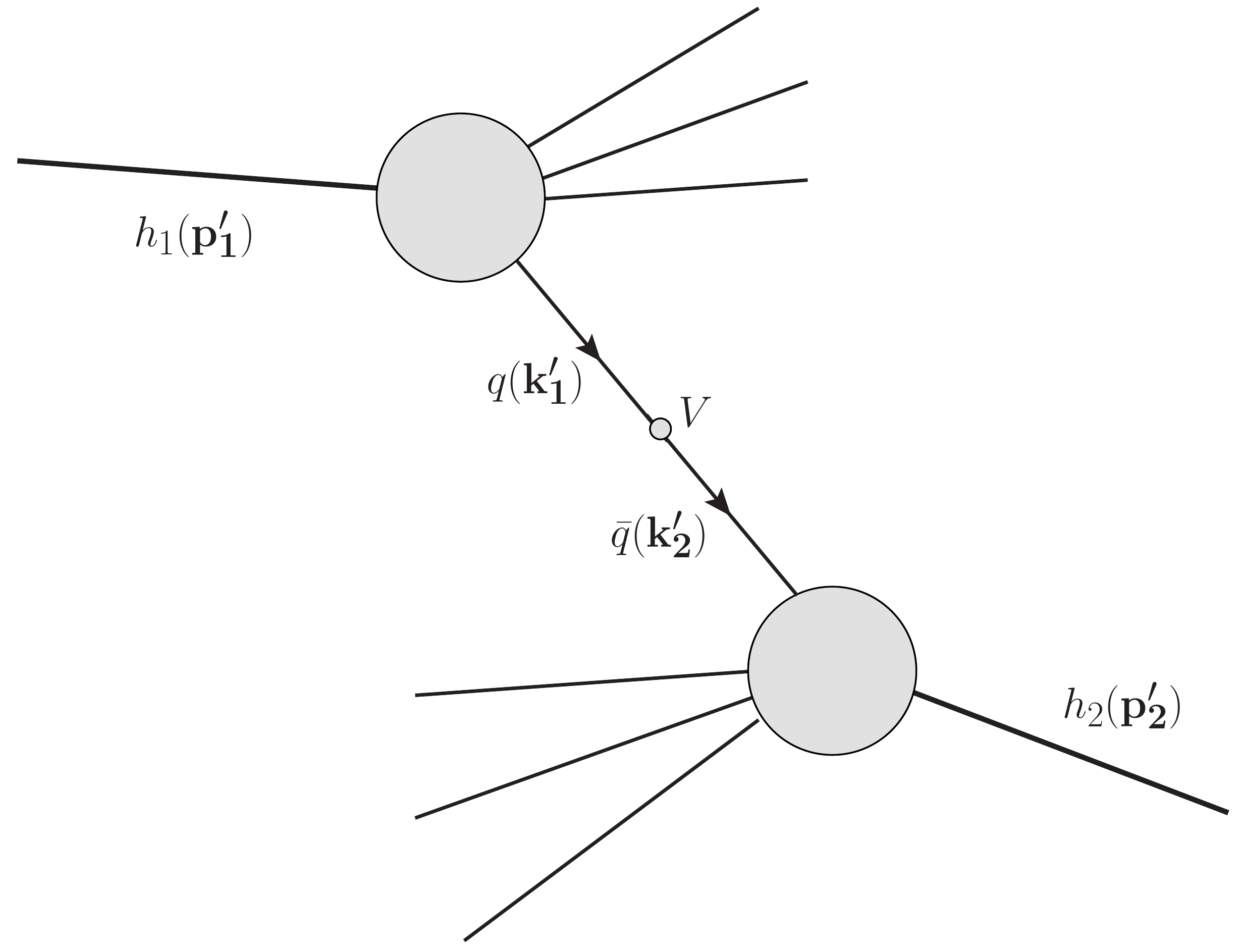}
\caption{The leading order Drell-Yan process \eqref{3.1} in the rest system of the produced 
vector boson $V=\gamma^*$ or $Z$.}
\label{SCD_fig6}
\end{center}
\end{figure}

In the rest system of the produced vector boson we have then the situation 
shown in figure \ref{SCD_fig6} for $V=\gamma^*$ or $Z$, a head-on collision of $q$ and ${\bar q}$, 
where
\be\label{3.5}
{\mathbf k}'_1+{\mathbf k}'_2=0.
\ee
Here and in the following momenta with a prime refer to the rest system of the $V$ boson. 
In leading order, this also is the c.m.\ system of the $q\bar q$ collision.  

\subsection{The general $q\bar q$ spin density matrix}
\label{The general spin density matrix}

In \cite{1,2} the ideas on the QCD vacuum structure, as sketched in section 2, were taken as a 
motivation to propose a more general framework for treating the Drell-Yan process than used at 
the time. 

The proposal was and is to allow for a general spin-density matrix for the $q\bar q$ system 
which includes all possible spin-momentum correlations. In addition, it was and is proposed to 
allow for correlations of the transverse momenta of $q$ and $\bar q$.

The latter correlations were motivated by effects of the chromomagnetic Lorentz force deflecting 
quark and antiquark in a correlated way in the background field. 
We should stress that this proposal was {\em motivated} by the QCD vacuum structure but was clearly 
supposed to be {\em considered in its own right}. 
We shall discuss below that the framework using transverse momentum dependent parton distribution 
functions (TMDs) is {\em included as a special case} in our more general framework which was proposed 
earlier. 

Thus, in \cite{1,2} the general two-particle density matrix for the $q\bar q$ system in the 
reaction \eqref{3.3} was analysed. 
In \cite{1} only partons $q$ and $\bar q$ collinear with $h_1$ and $h_2$, respectively, 
were considered and their spin and colour correlations analysed. 
It was shown in \cite{133} that colour correlations lead in general to infrared divergences 
in higher order calculations. 
This is not dramatic since the confinement radius provides, anyway, a natural infrared cutoff. 
Nonetheless, in \cite{2,3,4}, colour correlations were assumed to be absent. 
In \cite{2} the {\em most general correlated} $q\bar q$ spin density matrix for \eqref{3.3} was 
written down allowing also for arbitrary, even {\em correlated}, transverse momenta of $q$ and $\bar q$. 
For this purpose a three-dimensional vector basis in the $q\bar q$ c.m.\ system was constructed:
\be\label{3.6}
{\mathbf e}'_1=\frac{({\mathbf p}'_1+{\mathbf p}'_2)\times {\mathbf k}'_1}{|({\mathbf p}'_1+{\mathbf p}'_2)\times 
{\mathbf k}'_1|}~,~
{\mathbf e}'_3={\mathbf k}'_1/|{\mathbf k}'_1|~,~
{\mathbf e}'_2={\mathbf e}'_3\times {\mathbf e}'_1.
\ee
The ansatz for the most general $q\bar q$ density matrix of \cite{2} reads then 
\ba\label{3.7}
\rho^{(q,\bar q)}({\mathbf k}'_1,{\mathbf p}'_1,{\mathbf p}'_2)&=&
\big(\rho^{(q,\bar q)}_{\alpha\beta,\alpha'\beta'}({\mathbf k}'_1,{\mathbf p}'_1,{\mathbf p}'_2)\big)
\\
&=&\frac14\big\{ {\mathbbm 1}\otimes {\mathbbm 1}+F_j({\boldsymbol \sigma}\cdot {\mathbf e}'_j)\otimes {\mathbbm 1}
+G_j {\mathbbm 1}\otimes ({\boldsymbol \sigma}\cdot {\mathbf e}'_j)
+H_{ij}({\boldsymbol \sigma}\cdot {\mathbf e}'_i)\otimes(\boldsymbol \sigma\cdot {\mathbf e}'_j)\big\}. \nn
\ea
Here
\be\label{3.8}
{\mathbbm 1}\otimes {\mathbbm 1}=(\delta_{\alpha\alpha'}~~\delta_{\beta\beta'}),~~
{\boldsymbol \sigma}\otimes {\mathbbm 1}=({\boldsymbol \sigma}_{\alpha\alpha'}~~\delta_{\beta\beta'}), \text{ etc}. 
\ee
with $\alpha,\alpha'$ and $\beta,\beta'$ the quark and antiquark spin indices, respectively. 
The parameters of $\rho^{(q,\bar q)}$ are two vectors, $\mathbf F$ and $\mathbf G$, and a second 
rank tensor $(H_{ij})$ which all may depend on the momenta ${\mathbf k}'_1,{\mathbf p}'_1$, 
and ${\mathbf p}'_2$, 
\ba\label{3.9}
{\mathbf F}&=&F_j\mathbf e'_j={\mathbf F}({\mathbf k}'_1,{\mathbf p}'_1,{\mathbf p}'_2),\nn\\
{\mathbf G}&=&G_j{\mathbf e}'_j={\mathbf G}({\mathbf k}'_1,{\mathbf p}'_1,{\mathbf p}'_2),\nn\\
\underline{H}&=&H_{ij}{\mathbf e}'_i\otimes{\mathbf e}'_j
=\underline{H}({\mathbf k}'_1,{\mathbf p}'_1,{\mathbf p}'_2).
\ea
Parity (P) invariance of the strong interactions implies 
\be\label{3.10}
\rho^{(q,\bar q)}({\mathbf k}'_1,{\mathbf p}'_1,{\mathbf p}'_2)
=\rho^{(q,\bar q)}(-{\mathbf k}'_1,-{\mathbf p}'_1,-{\mathbf p}'_2).
\ee
Therefore, ${\mathbf F}$ and ${\mathbf G}$ must be pseudovectors, $\underline{H}$ a P-even tensor: 
\ba\label{3.11}
&&{\mathbf F}(-{\mathbf k}'_1,-{\mathbf p}'_1,-{\mathbf p}'_2)={\mathbf F}({\mathbf k}'_1,{\mathbf p}'_1,{\mathbf p}'_2),\nn\\
&&{\mathbf G}(-{\mathbf k}'_1,-{\mathbf p}'_1,-{\mathbf p}'_2)={\mathbf G}({\mathbf k}'_1,{\mathbf p}'_1,{\mathbf p}'_2),\nn\\
&&\underline{H}(-{\mathbf k}'_1,-{\mathbf p}'_1,-{\mathbf p}'_2)=\underline{H}({\mathbf k}'_1,{\mathbf p}'_1,{\mathbf p}'_2).
\ea
The standard assumption of unpolarised quarks and antiquarks corresponds to 
\be\label{3.12}
\begin{split}
&{\mathbf F}=0~,~{\mathbf G}=0~,~\underline{H}=0,\\
&\rho^{(q,{\bar q})}_{\text{standard}}=\frac14{\mathbbm 1}\otimes{\mathbbm 1}. 
\end{split}
\ee

At this point it is appropriate to discuss the dependences of the functions $F_j,G_j$, 
and $H_{ij}$ of \eqref{3.7} on other parameters than the momenta 
indicated explicitly in \eqref{3.9}. Already in \cite{1} it was written that the general 
density-matrix framework does {\em not} require the {\em same} non-perturbative effects for 
muon pair production in proton-proton and in other hadron-hadron collisions. 
Thus, in the general density-matrix formalism $\rho^{(\rho,\bar \rho)}$ should be allowed 
to depend on the nature of the parent hadrons $h_1$ and $h_2$. 
Let us also recall point (iii) of the summary in section \ref{The QCD vaccum structure}. 
The shielding of the chromomagnetic vacuum fields may be different in different hadrons and, 
thus, lead to a dependence of $\rho^{(q,\bar q)}$ on the hadrons $h_1$ and $h_2$. 
For given hadrons $h_1$ and $h_2$ $\rho^{(q,\bar q)}$ certainly should depend on the 
quark flavour $q$. This is clear from the whole discussion of the ``synchrotron'' effects of 
the chromomagnetic background fields in \cite{1,2,3,4} and section \ref{The QCD vaccum structure}. 
``Synchrotron'' effects clearly should be quite different for very light $u$ and $d$ quarks 
compared to heavier $s,c$, or $b$ quarks. Thus, in the general density-matrix 
approach $\rho^{(q,\bar q)}$ should be allowed to depend on the hadrons $h_1$ and $h_2$ as 
well as on the quark flavour $q$. 

In order to calculate in our general framework the cross section for the Drell-Yan processes 
\eqref{3.1} one has to evaluate the production matrix for $q+\bar q\to V(k,\epsilon)$ using the 
density matrix \eqref{3.7}. Then, the vector-boson production matrix from $q\bar q$ annihilation 
has to be summed over quark flavours and integrated over the momentum distributions of $q$ and $\bar q$, 
which in \cite{2} were also allowed to be correlated, to get the overall vector-boson production 
matrix for $h_1+h_2\to V+X$. 
Finally, this production matrix has to be contracted with the decay matrix for the decay of $V$ 
into the channel one wants to observe. Typically one considers leptonic decays $(l=e,\mu)$: 
\ba\label{3.13}
\gamma^*&\to& l^+l^-,\nn\\
Z&\to&l^+l^-
\ea
and similarly for $W$ production \eqref{3.2} the decays
\ba\label{3.14}
W^+&\to&l^+\nu_l,\nn\\
W^-&\to&l^-\bar\nu_l.
\ea
All formulae for performing this program are given in \cite{2}; see also appendix A of the present 
paper where a misprint of \cite{2} is corrected. 
Furthermore, in \cite{2} various predictions for the process \eqref{3.1} with $V=\gamma^*$ and $Z$ 
were worked out. 
At this point we want to point out that the work of \cite{2} was started as a natural continuation 
of \cite{1} and was half way completed without the authors knowing about the relevant experiments. 
Only then, in a discussion, H. J. Pirner kindly pointed out to the present author and his collaborators 
that in the NA10 experiment \cite{6,7} an extensive study of the lepton-pair angular distribution in 
the Drell-Yan process had been done. 
Of course, this was an exciting moment for us. 
Was everything in the experimental distributions according to the standard $q\bar q$ density matrix, 
\eqref{3.12}, or was there room for some non-standard effects? 
It turned out that there was indeed a {\em large} deviation from the standard expectation for the 
lepton-pair angular distribution as we shall recall in the next section. 

\subsection{Comparison with the NA10 data}
\label{Comparison with the the NA10 data}

In the NA10 experiment \cite{6,7} the Drell-Yan reaction in $\pi^-$ nucleon collisions was studied:
\ba\label{3.15}
\pi^-+N&\to&\gamma^*+X.\nn\\
&  & \drsh \mu^++\mu^-
\ea
The momenta of the incident $\pi^-$ were $140$ GeV$/c$, $194$ GeV$/c$, and $286$ GeV$/c$, the targets 
were deuterium and tungsten. The experiment collected enough statistics to make a detailed 
investigation of the muon's angular distributions. 

The lepton-pair distribution can be analysed in the so-called Collins-Soper (CS) frame where the 
following basis vectors are introduced in the $\gamma^*$ rest frame 
\ba\label{3.16}
\mathbf e'_{1,\text{CS}}&=&\frac{\mathbf{\hat p'_1}+\mathbf{\hat p'_2}}{|\mathbf{\hat p'_1}+\mathbf{\hat p'_2}|},\nn\\
\mathbf e'_{2,\text{CS}}&=&\frac{\mathbf{\hat p'_1}\times \mathbf{\hat p'_2}}{|\mathbf{\hat p'_1}\times\mathbf{\hat p'_2}|},\nn\\
\mathbf e'_{3,\text{CS}}&=&\frac{\mathbf{\hat p'_1}-\mathbf{\hat p'_2}}{|\mathbf{\hat p'_1}-\mathbf{\hat p'_2}|}.
\ea
Here $\mathbf p'_{1,2}$ are the momenta of $h_{1,2}$ in the $\gamma^*$ rest frame 
and $\mathbf{\hat p'_i}=\mathbf{p'_i}/ 
|\mathbf{p'_i}|,\:i=1,2$. The general formula for the angular distribution of the 
lepton $l^+$ in the Drell-Yan reaction \eqref{3.15} reads
\ba\label{3.17}
\frac 1\sigma \frac{d\sigma}{d\Omega'}=\frac{3}{4\pi}\frac{1}{\lambda+3}
\big(1+\lambda\cos^2\theta+\mu\sin(2\theta)\cos\phi+\frac\nu2\sin^2\theta\cos(2\phi)\big).
\ea
Here $\theta,\phi$ are the polar and azimuthal angles, respectively, of the $l^+$ momentum in the 
CS frame. 
The functions $\lambda,\mu$, and $\nu$ depend on the other kinematic variables, notably 
the pseudorapidity $\eta$, 
the absolute value of the transverse momentum, $|\mathbf{k}_T|$, and the mass of the virtual 
photon $\gamma^*$. 
We remark that there exists now a ``Trento convention 2012'' for defining the transformation 
from the c.m.\ system to the vector-boson rest system and for the angles $\theta,\phi$; see \cite{21}. 
For obvious reasons we stick here to the original conventions used in \cite{2}, but we give 
the relations to the new Trento convention in appendix B. 

\begin{figure}[ht]
\begin{center}
\includegraphics[width=0.45\textwidth]{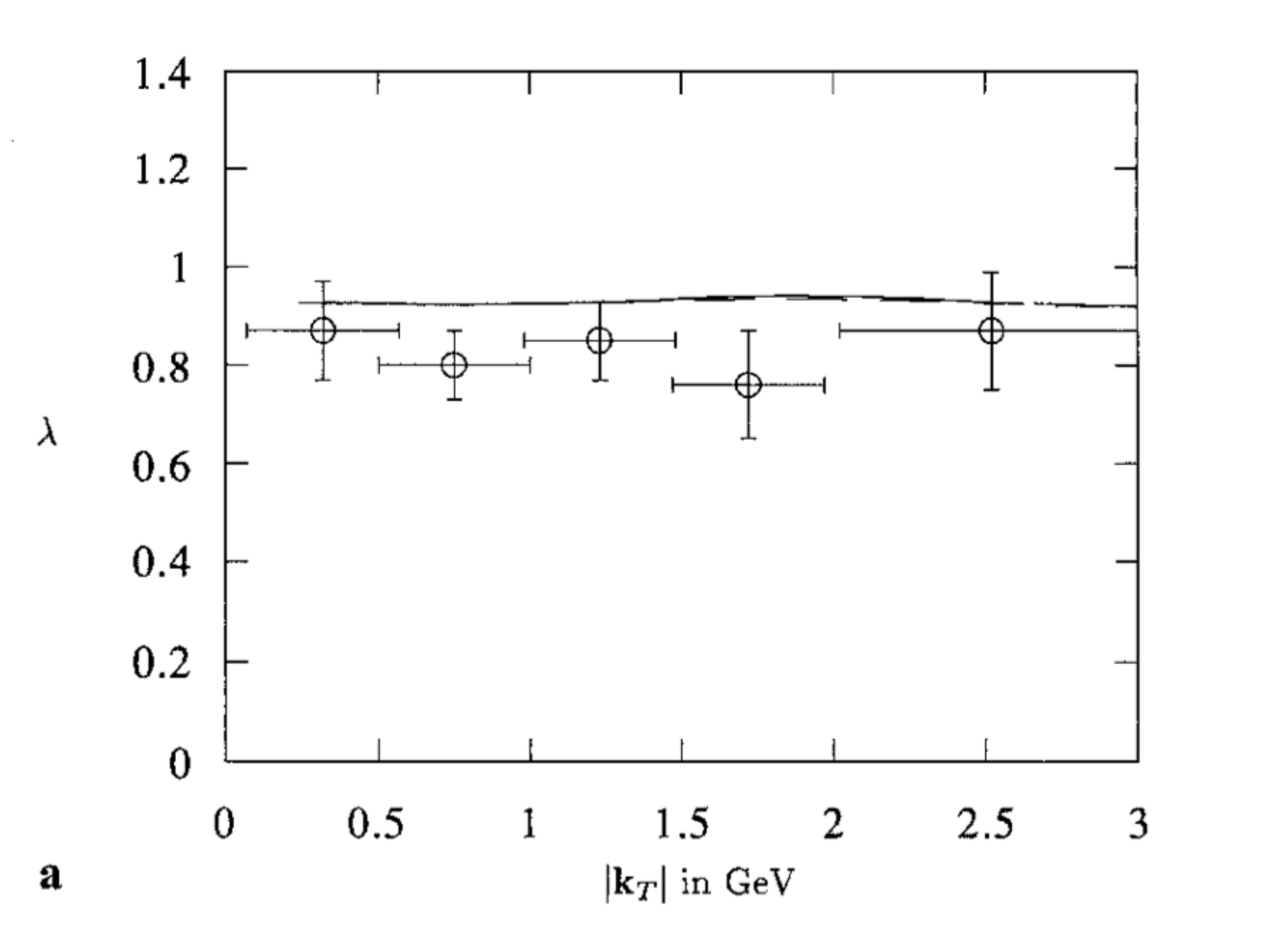}
\includegraphics[width=0.45\textwidth]{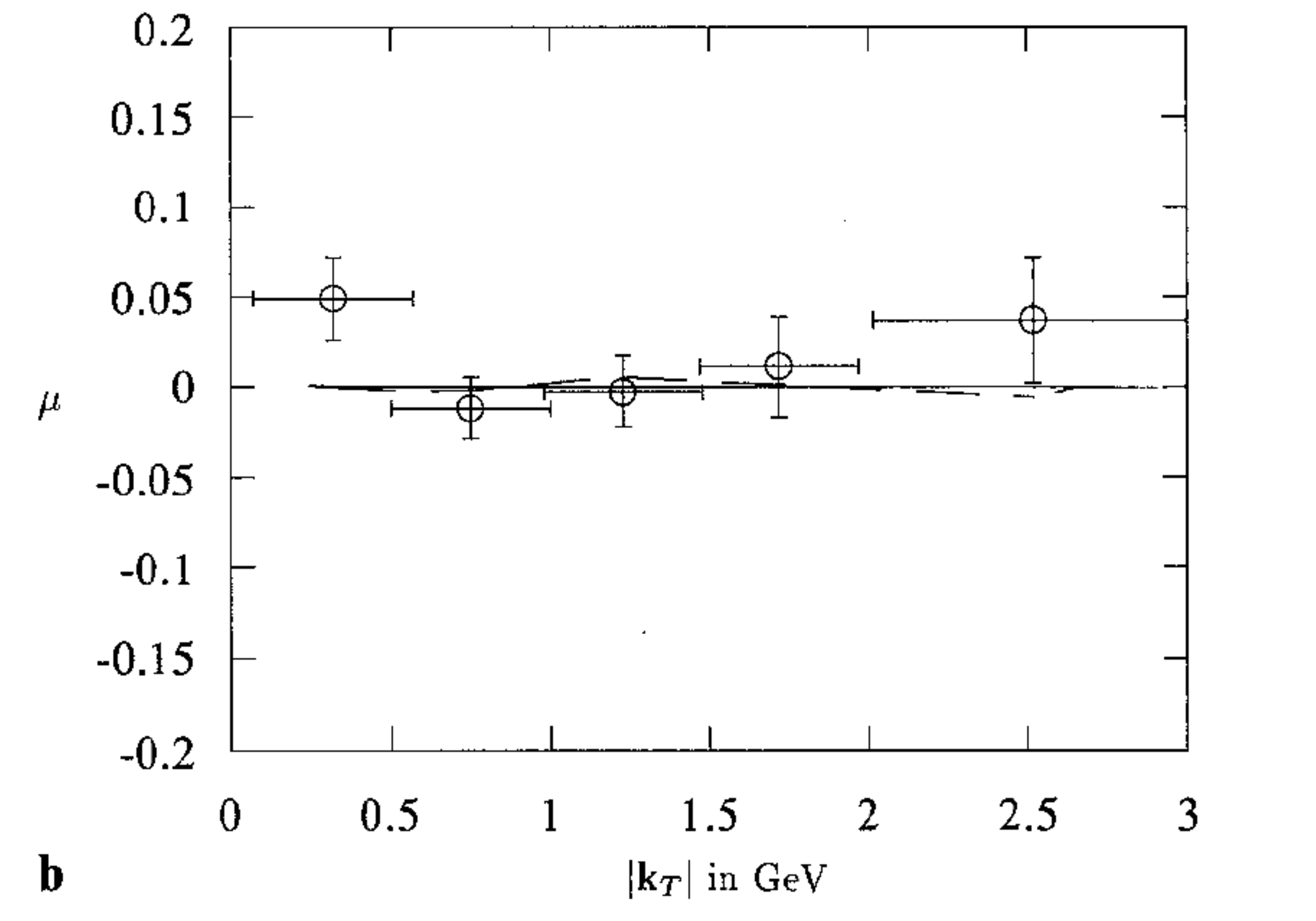}
\includegraphics[width=0.45\textwidth]{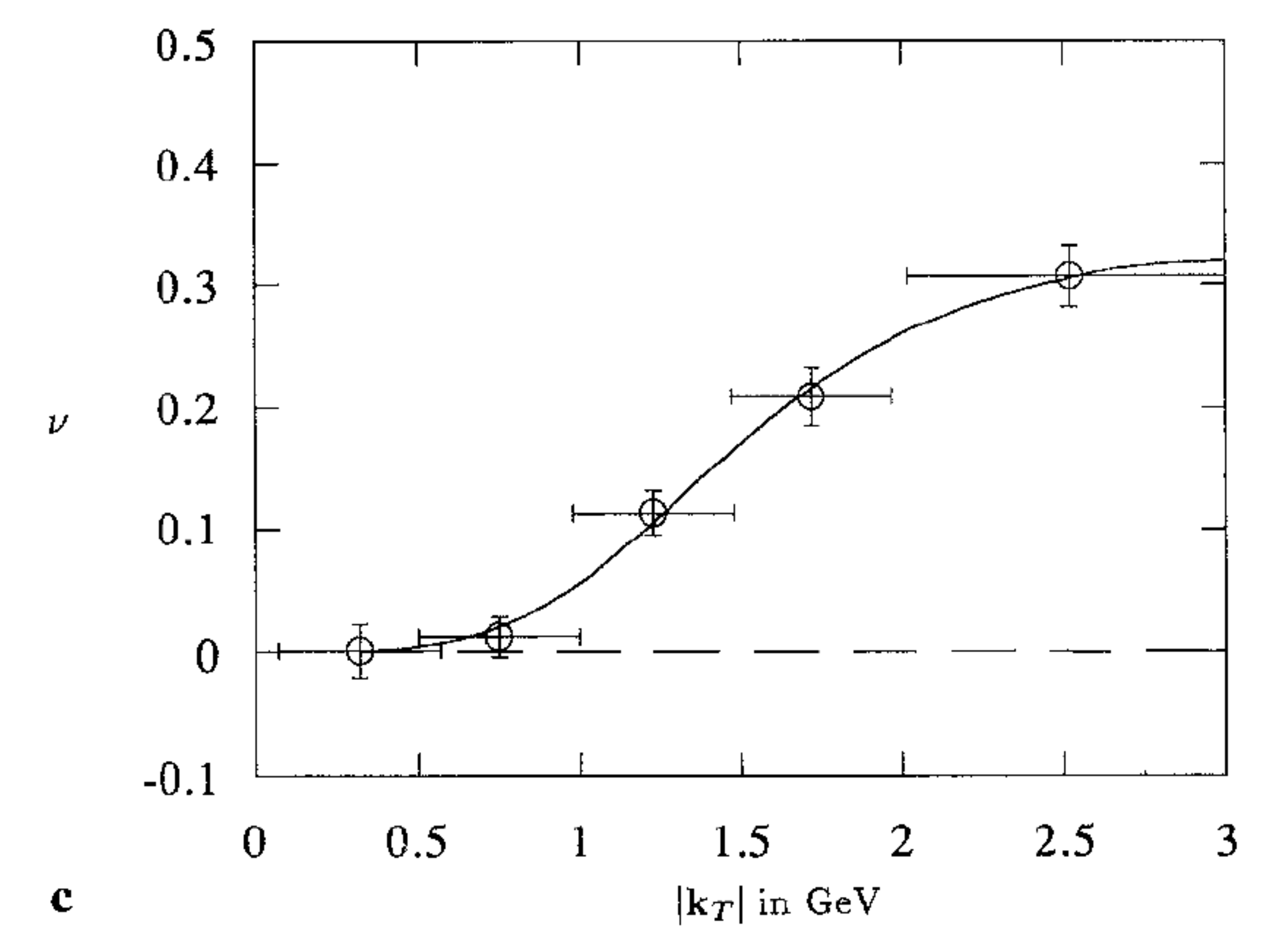}
\end{center}
\caption{The $|\mathbf k_T|$ dependence of the structure functions $\lambda,\mu,\nu$ for $\pi^-N$ 
collisions at $194$ GeV$/c$ incident $\pi^-$ momentum. 
The pseudorapidity and mass of $\gamma^*$ are $\eta=0$ and $m_{\gamma^*}=8$ GeV, respectively. 
The data are from Table 4 of \cite{7}. The dashed lines are for the standard $q\bar q$ density 
matrix \eqref{3.12}, 
the solid lines for the non-trivial $q\bar q$ density matrix \eqref{3.7} with $\kappa$ 
from \eqref{3.23}, \eqref{3.24}.
See the discussion there for the details. This figure is reproduced from figure 8 of \cite{2}.
\label{SCD_fig7}}
\end{figure}

We return to the discussion of the NA10 experiment. With the standard $q\bar q$ density 
matrix \eqref{3.12} one finds the Lam-Tung relation 
\be\label{3.18}
1-\lambda-2\nu=0
\ee
which is valid also to order $\alpha_s$; see\cite{134}. In \cite{135} it was found that even 
to order $\alpha^2_s$ the relation \eqref{3.18} is hardly changed with the standard density matrix. 
But in the experiment \cite{6,7} a {\em large} violation of the Lam-Tung relation \eqref{3.18} is 
found; see figure \ref{SCD_fig7} which is taken from figure 8 of \cite{2}. 
We see that in the $|\mathbf{k}_T|$ interval explored by the experiment $\lambda\approx 1,\mu\approx 0$. 
Then, the Lam-Tung relation \eqref{3.18} predicts:
\be\label{3.19}
\nu\approx 0
\ee
in violent disagreement with experiment. The theory with the standard $q\bar q$ density matrix 
\eqref{3.12} gives the dashed lines in figure \ref{SCD_fig7}, in accordance with \eqref{3.19}. 
Also soft gluon resummations do not change this result appreciably; see \cite{136}. What can one 
say on the data from figure \ref{SCD_fig7} assuming a non-trivial $q\bar q$ density matrix \eqref{3.7}? 

To answer this question it is useful to discuss first which elements of the density matrix 
\eqref{3.7} can be probed in the Drell-Yan reaction \eqref{3.1} with $V=\gamma^*$ and $Z$ at 
leading order. 
We consider the $q\bar q$ annihilation \eqref{3.3} for massless quarks. 
Then $\gamma_5$ invariance of the $Vq\bar q$ vertex tells us that the annihilation can only 
occur in the following helicity configurations
\be\label{3.20}
q_L\bar q_R\to V~,~q_R\bar q_L\to V 
\ee
where $L$ and $R$ stand for left- and right-handed polarisations, respectively. 
With the standard representation of the Pauli matrices and using the coordinate 
system \eqref{3.6} the $2$-spinors of $q$ and $\bar q$ of definite helicity are as follows:
\ba\label{3.21}
\text{quark }q:\qquad && \chi_R=\left(\begin{array}{c}1\\0\end{array}\right)~,~
\chi_L=\left(\begin{array}{c}0\\1\end{array}\right),
\\
\label{3.22}
\text{antiquark }\bar q:\qquad &&
\chi'_R=
\left(\begin{array}{c}0\\1\end{array}\right)~,~
\chi'_L=\left(\begin{array}{c}1\\0\end{array}\right).
\ea
From this we get the density matrix $\rho^{(q,\,\bar q)}$ \eqref{3.7} in the helicity basis as 
shown in table \ref{Tab:2}. 
With \eqref{3.20} we see that only the matrix elements of the $2\times 2$ submatrix corresponding 
to the entries $RL$ and $LR$ enter in the Drell-Yan reaction \eqref{3.1} 
with $V=\gamma^*$ or $Z$. Furthermore, for the ordinary Drell-Yan reaction, $V=\gamma^*$, 
with $\gamma^*\to l^+l^-$ and no observation of the lepton polarisations only 
$H_{33}$, $H_{11}-H_{22}$, and $H_{12} + H_{21}$ remain as relevant parameters; see appendix A. 
It turns out that $1+H_{33}$ enters in the calculation of the overall rate. 
In \cite{2} $H_{12} + H_{21}$ was set to zero and it was found that then the parameter 
\be\label{3.23}
\kappa = \frac{H_{22}-H_{11}}{1+H_{33}}
\ee
determines the lepton angular distribution. In \cite{2} an {\em ansatz} was made for ${\kappa}$ at 
pseudorapidity $\eta=0$:
\be\label{3.24}
\kappa = {\kappa}_0\frac{|\mathbf k_T|^4}{|\mathbf k_T|^4+m^4_T}
\ee
with ${\kappa}_0$ and $m_T$ as parameters. It was found that instead of the Lam-Tung relation 
\eqref{3.18} one has now 
\be\label{3.25}
1-\lambda-2\nu\approx-4{\kappa}.
\ee

\begin{table}[tbp]
\caption{The matrix elements $ 4\langle q_A,\bar q_B |\rho^{(q,\bar q)}|q_{A'},\bar q_{B'}\rangle $
where $\rho^{(q,\,\bar q)}$ is given in \eqref{3.7} and $A,B,A',B'\in \{R,L\}$. Only the 
upper-left 2$\times$2 submatrix is relevant for the Drell-Yan reaction \eqref{3.1}.
}
\scriptsize
\begin{tabular}{r|c|c|c|c|}
$\begin{array}{c}
~~~~~A'B'\\
\raisebox{0pt}[0pt][0pt]{
{$\cdot$\raisebox{-0.3ex}
{$\cdot$}\raisebox{-0.7ex}
{$\cdot$}\raisebox{-1.1ex}
{$\cdot$}\raisebox{-1.5ex}
{$\cdot$}\raisebox{-1.9ex}
{$\cdot$}\raisebox{-2.3ex}
{$\cdot$}\raisebox{-2.7ex}
{$\cdot$}\raisebox{-3.1ex}
{$\cdot$}}}
\\ \\AB~~~~~
\end{array}$
&$RL$&$LR$&$RR$&$LL$\\ \hline
&&&&\\
$RL$~~~~~~&$1+F_3+G_3+H_{33}$&$H_{11}-H_{22}-iH_{12}-iH_{21}$&
$G_1-iG_2+H_{31}-iH_{32}$&
$F_1-iF_2+H_{13}-iH_{23}$\\
&&&&\\ \hline 
&&&&\\
$LR$~~~~~~&$H_{11}-H_{22}+iH_{12}+iH_{21}$&
$1-F_3-G_3+H_{33}$&
$F_1+iF_2-H_{13}-iH_{23}$&
$G_1+iG_2-H_{31}-iH_{32}$\\
&&&& \\ \hline 
&&&& \\
$RR$~~~~~~&$G_1+iG_2+H_{31}+iH_{32}$&$F_1-iF_2-H_{13}+iH_{23}$&
$1+F_3-G_3-H_{33}$&$H_{11}+H_{22}+iH_{12}-iH_{21}$\\
&&&& \\\hline
&&&&\\
$LL$~~~~~~&$F_1+iF_2+H_{13}+iH_{23}$&
$G_1-iG_2-H_{31}+iH_{32}$&
$H_{11}+H_{22}-iH_{12}+iH_{21}$&
$1-F_3+G_3-H_{33}$\\
&&&& \\ \hline
\end{tabular}
\label{Tab:2}
\end{table}

\normalsize

The complete calculation for $\lambda,\mu$, and $\nu$ has to take into account the integration over 
the parton distributions. The results are shown in figure \ref{SCD_fig7} as the solid lines for the choice 
\ba\label{3.26}
{\kappa}_0&=&0.17,\nn\\
m_T&=&1.5 \text{ GeV}
\ea
in \eqref{3.24}. The agreement with experiment is quite satisfactory. In \cite{2} it was also shown 
that the sign of ${\kappa}$ is precisely as expected from the chromomagnetic Sokolov-Ternov effect. 
Indeed, consider figure \ref{SCD_fig3} with $q$ and $\bar q$ annihilating in the common background 
field $\mathbf B^a$. 
Then, $q$ and $\bar q$ must come with corresponding colour and anticolour, and thus, will get opposite 
transverse polarisation due to the chromomagnetic Sokolov-Ternov effect. 
Taking the reaction plane as spanned by $\mathbf k'_1$ and $\mathbf p'_1+\mathbf p'_2$, see 
figure \ref{SCD_fig6} and \eqref{3.6}, the transverse direction is spanned by $\mathbf e'_1$. 
In \cite{2} such a correlated transverse polarisation of $q$ and $\bar q$ was found to lead to 
a density matrix \eqref{3.7} with 
\be\label{3.27}
F_1=s~,~G_1=-s~,~H_{11}=-s^2,
\ee
and all other elements $F_j,G_j$ and $H_{ij}$ equal to zero. Here $s$, with $|s|\leq 1$, is 
the degree of transverse polarisation of the quark. Clearly, from \eqref{3.23} and \eqref{3.27} 
we get ${\kappa}\geq 0$. 
We emphasise that \eqref{3.27} is only supposed to give an {\em example} of parameters leading to 
a density matrix with ${\kappa}\geq 0$. It is {\em not} to be cited as ``the prediction of our model''. 
We should also emphasise that the simple ansatz for ${\kappa}$ in \eqref{3.24} made in comparison 
with the data shown in figure 7, was {\em never} supposed to be universally valid. On the contrary, 
after (3.13) of \cite{2} it was written  that the parameters $\mathbf F,\mathbf G$ and $\underline{H}$ 
of the density matrix $\rho^{(q,\bar q)}$ are real functions of the invariants of the problem. 
We shall come back to this point in section \ref{s3.5} and appendix A below.

\subsection{Relation to the TMD approach}
\label{Relation to the TMD approach}
When we discussed at the time the results of the paper \cite{2}, as outlined in 
section \ref{Comparison with the the NA10 data} here, with theorists and experimentalists 
there was little resonance. 
One reason may be that at the time no further experiments studying the Drell-Yan reaction were 
on the horizon. 
Thus, the present author went on to work on other topics. His interest in the Drell-Yan problem 
was rekindled at a physics meeting in 2003 where he met D.~Boer. 
The latter told him about the work on this problem he had done \cite{137}. In the common 
paper \cite{5} the relation of the respective approaches was discussed. 
The conclusions were as follows. In the TMD approach of \cite{137} a {\em factorising} density 
matrix is assumed for the $q\bar q$ pair: 
\be\label{3.28}
\rho^{(q,\bar q)}(\mathbf k'_1,\mathbf p'_1,\mathbf p'_2)=
\rho^{(q)}(\mathbf k'_1,\mathbf p'_1)\otimes \rho^{(\bar q)}(-\mathbf k'_1,\mathbf p'_2).
\ee
Here the density matrix for the quark from hadron $h_1$ is allowed to depend only on the quark 
momentum $\mathbf k'_1$ and the $h_1$ momentum $\mathbf p'_1$. 
Similarly, $\rho^{(\bar q)}$ depends here only on the momenta of the antiquark $(-\mathbf k'_1)$ and 
of hadron $h_2$ $(\mathbf p'_2)$. Thus, we get
\ba\label{3.29}
&\rho^{(q)}(\mathbf k'_1,\mathbf p'_1)=&
\frac12 \big\{ {\mathbbm 1}+\mathbf F^B(\mathbf k'_1,\mathbf p'_1)\cdot \boldsymbol \sigma\big\},\nn\\
&\rho^{(\bar q)}(-\mathbf k'_1,\mathbf p'_2)=&
\frac12\left\{ {\mathbbm 1}+\mathbf G^B (-\mathbf k'_1,\mathbf p'_2)\cdot \boldsymbol \sigma\right\}.
\ea
From P invariance of the strong interactions $\mathbf F^B$ and $\mathbf G^B$ must be 
pseudovectors, see \eqref{3.11}, and therefore we must have 
\begin{align}
\label{3.30}
\mathbf F^B(\mathbf k'_1,\mathbf p'_1) \;\, &\propto \;\, \mathbf k'_1\times \mathbf p'_1 \;\,\propto \;\, \mathbf e'_3 \times \mathbf p'_1,&\nn\\
\mathbf G^B(-\mathbf k'_1,\mathbf p'_2) \;\,&\propto \;\, (-\mathbf k'_1)\times \mathbf p'_2  \;\,\propto \;\, -\mathbf e'_3 \times \mathbf p'_2.&
\end{align}

With the functions $f_1,h^\perp_1,\bar f_1,\bar h^\perp_1$ as defined in \cite{137,5} 
the ansatz for $\rho^{(q,\bar q)}$ using TMDs finally reads as in \eqref{3.7}, \eqref{3.9} 
with the replacements
\ba\label{3.31}
\mathbf F(\mathbf k'_1,\mathbf p'_1,\mathbf p'_2)\to
\mathbf F^B(\mathbf k'_1,\mathbf p'_1)&=&
\frac{h^\perp_1x_1}{f_1M_1}\left(\mathbf e'_3\times \mathbf p'_1\right),\\
\mathbf G(\mathbf k'_1,\mathbf p'_1,\mathbf p'_2)\to\mathbf G^B(-\mathbf k'_1,\mathbf p'_2)
&=&-\frac{\bar h^\perp_1x_2}{\bar f_1 M_2}(\mathbf e'_3\times\mathbf p'_2),\label{3.32}\\
H_{ij}(\mathbf k'_1,\mathbf p'_1,\mathbf p'_2)\to 
H^B_{ij}(\mathbf k'_1,\mathbf p'_1,\mathbf p'_2)&=&
F^B_i(\mathbf k'_1,\mathbf p'_1)G^B_j(-\mathbf k'_1,\mathbf p'_2).\label{3.33}
\ea
Here $f_1,\dots ,\bar h^\perp_1$ have a functional dependence as follows:
\ba\label{3.34}
f_1&=&f_1(x_1,\mathbf k^2_{1T}),\nn\\
h^\perp_1&=&h^\perp_1(x_1,\mathbf k^2_{1T}),\nn\\
\bar f_1&=&\bar f_1(x_2,\mathbf k^2_{2T}),\nn\\
\bar h^\perp_1&=&\bar h^\perp_1(x_2,\mathbf k^2_{2T})
\ea 
where $x_i$ and $\mathbf k^2_{iT}$ are defined in \eqref{3.4}.

We see from \eqref{3.30} to \eqref{3.33} that in the TMD approach we have, in particular,
\ba\label{3.35a}
F_3^B &=& G_3^B = 0, \nn\\
H_{33}^B &=& 0.
\ea
Clearly, the TMD framework where only factorising $q\bar q$ density matrices are considered 
is {\em included as a} {\em special case} in our framework of  \eqref{3.7} to \eqref{3.11}. 
Thus, it makes {\em no sense} to ask for an effect which is describable by the TMD approach 
and not in our more general framework. 
But it makes a lot of sense to ask if there are observable effects which {\em cannot} 
be described by a factorising $q\bar q$ density matrix, \eqref{3.28} to \eqref{3.33}, 
but require a general $q\bar q$ density matrix. Such effects would point to the phenomenon of 
{\em parton entanglement}.

At this point it is appropriate to clarify a misunderstanding with which the present author 
frequently is confronted by colleagues. In \cite{2} it was proposed to use a {\em non-trivial} spin-density matrix 
for the $q\bar q$ system; see \eqref{3.7}. If this density matrix would be factorising or entangled was, at the 
time of writing \cite{2}, left open and not at the forefront of the considerations. In fact, an explicit 
example of a non-trivial density matrix discussed in (3.22) ff. of \cite{2} and recalled in \eqref{3.27} of the 
present paper is of the factorising form concerning the spins:
\be\label{3.35b}
\rho^{(q,\bar q)}(\mathbf k'_1,\mathbf p'_1,\mathbf p'_2)=
\frac12 \big\{ {\mathbbm 1}+\mathbf F\cdot \boldsymbol \sigma\big\}\otimes
\frac12\left\{ {\mathbbm 1}+\mathbf G\cdot \boldsymbol \sigma\right\}
\ee
where
\begin{eqnarray}
\mathbf{F} =& s\mathbf{e'_1} \;\, \propto \;\, (\mathbf{p'_1} + \mathbf{p'_2})\times \mathbf k'_1,\nonumber\\
\mathbf{G} =& -s\mathbf{e'_1} \;\, \propto \;\, (\mathbf{p'_1} + \mathbf{p'_2})\times (\mathbf{- k'_1}).\nonumber\\
\label{3.35c}
\end{eqnarray}
Clearly \eqref{3.35b} is quite close to the TMD ansatz \eqref{3.28} to \eqref{3.30}. The difference is that 
in \eqref{3.35b}, \eqref{3.35c} $\mathbf{F}$ and $\mathbf{G}$ depend on the sum of the two vectors $\mathbf{p'_1}$ and $\mathbf{p'_2}$ 
whereas in \eqref{3.30} $\mathbf{F^B}$ depends only on $\mathbf{p'_1}$ and $\mathbf{G^B}$ only on $\mathbf{p'_2}$ 
in addition to $\mathbf{k'_1}$. For the present author the question of parton entanglement only came to 
the forefront during the work on ref. \cite{5}. Entanglement is {\em not} describable in the TMD framework 
but there is no problem 
to describe and parametrise it in our more general framework. We shall, therefore, discuss in the next section possible 
signatures of parton entanglement.

\subsection{Signatures of parton entanglement in the Drell-Yan process}
\label{Signatures of parton entanglement}
Here we discuss some effects which, if observed, would point towards a general, non-factorising, 
$q\bar q$ density matrix \eqref{3.7}.\\

\noindent
{\bf Correlations of transverse momenta of quark and antiquark}\\[2mm]
In our leading order calculation we get for the mean transverse momentum of the vector 
boson $V=\gamma^*,\,Z$ in the reaction \eqref{1.1}
\ba \label{3.35}
\langle\mathbf{k}^2_T\rangle &=& \langle\left(\mbf{k}_{1T}+\mbf{k}_{2T}\right)^2\rangle \nn \\
&=& \langle\mathbf{k}^2_{1T}\rangle+\langle\mathbf{k}^2_{2T}\rangle+2\langle\mathbf{k}_{1T}\cdot \mathbf{k}_{2T}  \rangle \: ;
\ea
see \eqref{3.3}, \eqref{3.4}.
Here $\mbf{k}_{1T}$ and $\mbv{k}{2T}$ refer to the quark $q$ and antiquark $\bar q$ 
transverse momenta, respectively, and the average is also over quark from $h_1$, antiquark 
from $h_2$ and vice versa.
If now the $q$ and $\bar q$  transverse momenta are uncorrelated we get:
\ba \label{3.36}
\langle \mbv{k}{1T}\cdot \mbv{k}{2T}\rangle &=& 0, \nn \\
\langle \mbv{k}{T}^2\rangle &=& \langle \mbv{k}{1T}^2\rangle + \langle \mathbf{k}_{2T}^2 \rangle.
\ea
On the other hand, maximal positive $k_T$ correlations imply
\ba \label{3.37}
\langle \mbv{k}{1T}\cdot \mbv{k}{2T}\rangle &=& \sqrt{\langle \mbv{k}{1T}^2\rangle} \sqrt{\langle \mbv{k}{2T}^2\rangle}, \nn \\
\langle \mbv{k}{T}^2\rangle &=& \left( \sqrt{\langle \mbv{k}{1T}^2\rangle} + \sqrt{\langle \mbv{k}{2T}^2\rangle} \right)^2 .
\ea
Suppose, as an example, that 
\be \label{3.38}
\langle \mbv{k}{1T}^2\rangle = \langle \mbv{k}{2T}^2\rangle .
\ee
Then, assuming no $k_T$ correlations, we get from \eqref{3.36}
\be \label{3.39}
\langle \mbv{k}{1T}^2\rangle = \frac{1}{2} \langle \mbv{k}{T}^2\rangle .
\ee
But with maximal $k_T$ correlations we find from \eqref{3.37}
\be \label{3.40}
\langle \mbv{k}{1T}^2\rangle = \frac{1}{4} \langle \mbv{k}{T}^2\rangle .
\ee
Already in \cite{2} an explicit distribution of the $q$ and $\bar q$ transverse momenta was 
given which interpolates between the two extreme cases above.
This is recalled and discussed in detail in appendix A.

The message from these considerations is the following. If positive $k_T$ correlations are present 
in the DY reaction these \emph{must} be taken into account 
when estimating the $q$ and $\bar q$ transverse momenta from the $k_T$ of the vector boson.
Neglecting the correlations the estimates of 
$\langle \mbv{k}{1T}^2 \rangle$, $\langle \mbv{k}{2T}^2 \rangle$ from DY will, for positive 
correlations,  be too large compared, for instance,
to estimates from semi-inclusive-deep-inelastic scattering (SIDIS).

It is interesting to note that, indeed, there is a tension between the $k_T$ determinations 
of $q$ and $\bar q$ from DY and SIDIS; see \cite{21}. This could point to parton entanglement 
as discussed here.
But one must be careful since, in reality, the influence of higher order QCD effects, 
both for DY and SIDIS, has to be investigated before one can draw more firm conclusions.\\

\noindent
{\bf Absolute normalisation of the DY cross section}\\[2mm]
We see from the discussion in \cite{2,5} and from \eqref{A.17b}, \eqref{A.25} and \eqref{A.35}, 
\eqref{A.36} that the absolute normalisation of the DY cross section is sensitive to ($1+H_{33}$). 
In the TMD approach one has strictly $H_{33}=0$; see \eqref{3.35a}. Thus, a good measurement of 
the absolute normalisation in the DY reaction could reveal parton entanglement.
Here the proton-antiproton reaction ($h_1 = p, \, h_2=\bar p$ in \eqref{1.1}) would be most 
suitable since there the parton distribution functions are well known.
The effect from $1+H_{33}$ is in essence the effect already discussed in \cite{122} on the basis 
of an instanton calculation.\\

\noindent
{\bf Angular distribution of the lepton pair in the reaction \eqref{1.1}}\\[2mm]
Here the task is, in principle, to determine all parameters 
$\mbv{F}{}, \mbv{G}{} \textrm{ and } H_{ij}$ in the general density matrix \eqref{3.7}.
However, we see from \eqref{A.17} and \eqref{A.17b} that from \eqref{1.1} 
with $V=\gamma^*$ and $Z$ only ($1+H_{33}$), ($F_3+G_3$), ($H_{11}-H_{22}$) and ($H_{12}+H_{21}$) 
can be determined.
Nonzero values for $(H_{11}-H_{22})$ and ($H_{12}+H_{21}$) are also allowed in the TMD approach, 
but there ($F_3+G_3$) and $H_{33}$ must be zero; see \eqref{3.35a}.
Thus, an experimental result of $F_3+G_3 \neq 0$ would point to parton entanglement. 
But ($F_3+G_3$) is hard to observe; see appendix A.
One may have a chance in $Z$ production; see \eqref{A.17}, \eqref{A.31}.
For the DY reaction \eqref{1.1} with  $V=\gamma^*$ one would need observation of 
the lepton polarisation. This seems very difficult. 
Maybe, the DY reaction with $\tau$ leptons could be suitable. Instanton effects for the angular distributions 
in the DY reaction were discussed in \cite{5,BRU}.

\subsection{Remarks on some recent experiments}\label{s3.5}
\label{Remarks on some recent experiments}

In this section we want to make comments on the findings of some recent experiments.

In the experiments \cite{14,15} on the DY reactions
\ba
p + d &\to& \gamma^* + X, \nn \\
p + p &\to& \gamma^* +X,
\ea
only very small, if any, non-standard QCD effects as discussed here were observed.
Does this present a problem to the views and the ans\"atze discussed in \cite{1,2,3,4} and the 
present article?
\emph{Certainly not!} In our general approach to the $q\bar q$ density matrix its parameters are 
left free; see section \ref{The general spin density matrix} and appendix A.
In our view, in a phenomenological approach, these parameters from non-perturbative QCD are to be 
determined from experimental data as is the case for the parton distribution functions.
We also recall point (iii) of the physical picture explained in the summary of 
section \ref{The QCD vaccum structure}.
It is quite probable that the shielding of vacuum effects due to soft gluons is more 
important for sea quarks in the nucleons than for valence quarks and antiquarks in pions 
and valence quarks in nucleons.
That is, the parameters of $\rho^{(q,\bar q)}$ \eqref{3.7} must be allowed to depend on the 
quark flavour $q$, on the type of hadrons $h_1,\,h_2$ in \eqref{1.1}, and on the kinematic variables;
see the discussion in appendix A.

In the experiment \cite{17} $e^+e^-$ pairs in the $Z$ mass region were studied in $p\bar p$ collisions 
at $\sqrt{s}=1.96 \text{ TeV}$.
Again, no significant non-standard QCD effects as discussed by us here were found.
Here we have to remark the following.
\begin{enumerate}[(1)]
 \item In the experiment \cite{17} the transverse momenta of the $e^+e^-$ pair are on average 
 quite large.
 Thus, our considerations, which here are restricted to the leading order process \eqref{3.3}, 
 are not directly applicable.
 A careful study of higher order QCD effects together with our non-standard ansatz would be necessary.
 \item The experiment \cite{17} does not distinguish $\gamma^*$ and $Z$ production. 
 As already emphasised in \cite{2} the non-standard effects from the parameters $(H_{11}-H_{22})$ 
 and $(H_{12}+H_{21})$
 in the $q\bar q$ density matrix \eqref{3.7} enter with opposite sign for $\gamma^*$ and $Z$ production.
 We see this by comparing \eqref{A.17} and \eqref{A.17b}.
 For $Z$ production, \eqref{A.17}, the term with $(H_{11}-H_{22})$ and $(H_{12}+H_{21})$ is 
 multiplied by $-(g_{Vq}^2-g^2_{Aq})$, for $\gamma^*$ production, \eqref{A.17b}, by $(-1)$.
 We note that for $\sin^2 \theta_W \approx 0.23$ we have from \eqref{A.7a}, \eqref{A.7b}
 \be\label{3.42}
-(g_{Vq}^2-g^2_{Aq}) > 0, \;\,\text{for } q=u,d.
 \ee
Thus, in a mixture of $\gamma^*$ and $Z$ events the effects of $(H_{11}-H_{22})$ and $(H_{12}+H_{21})$ 
will be reduced.
 \item At these high energies, $\sqrt{s}=1.96 \text{ TeV}$, $q\bar q$ annihilation for quarks 
 other than $q=u, d$ will be non-negligible.
 Already for $s$ quarks we expect smaller and for $c$ and $b$ quarks even much smaller 
 non-standard QCD effects than for $u$ and $d$ quarks; see the discussion in 
 section \ref{The QCD vaccum structure} and in \cite{1,2,3,4}.
\end{enumerate}
To summarise: the interesting findings of the experiments \cite{14,15} put constraints on 
the parameters of $\rho^{(q,\bar q)}$ for $pp$ and $pd$ collisions. 
For experiment \cite{17} at least an analysis of $\gam$ and $Z$ production 
separately -- and also of the $\gam$-$Z$ interference term -- would be necessary before 
one could draw further conclusions on $\rho^{(q,\bar q)}$.

\section{Higgs-boson production and parton entanglement}
\label{Higgs-boson production}

\begin{figure}[htb]
\begin{center}
\includegraphics[width=0.65\textwidth]{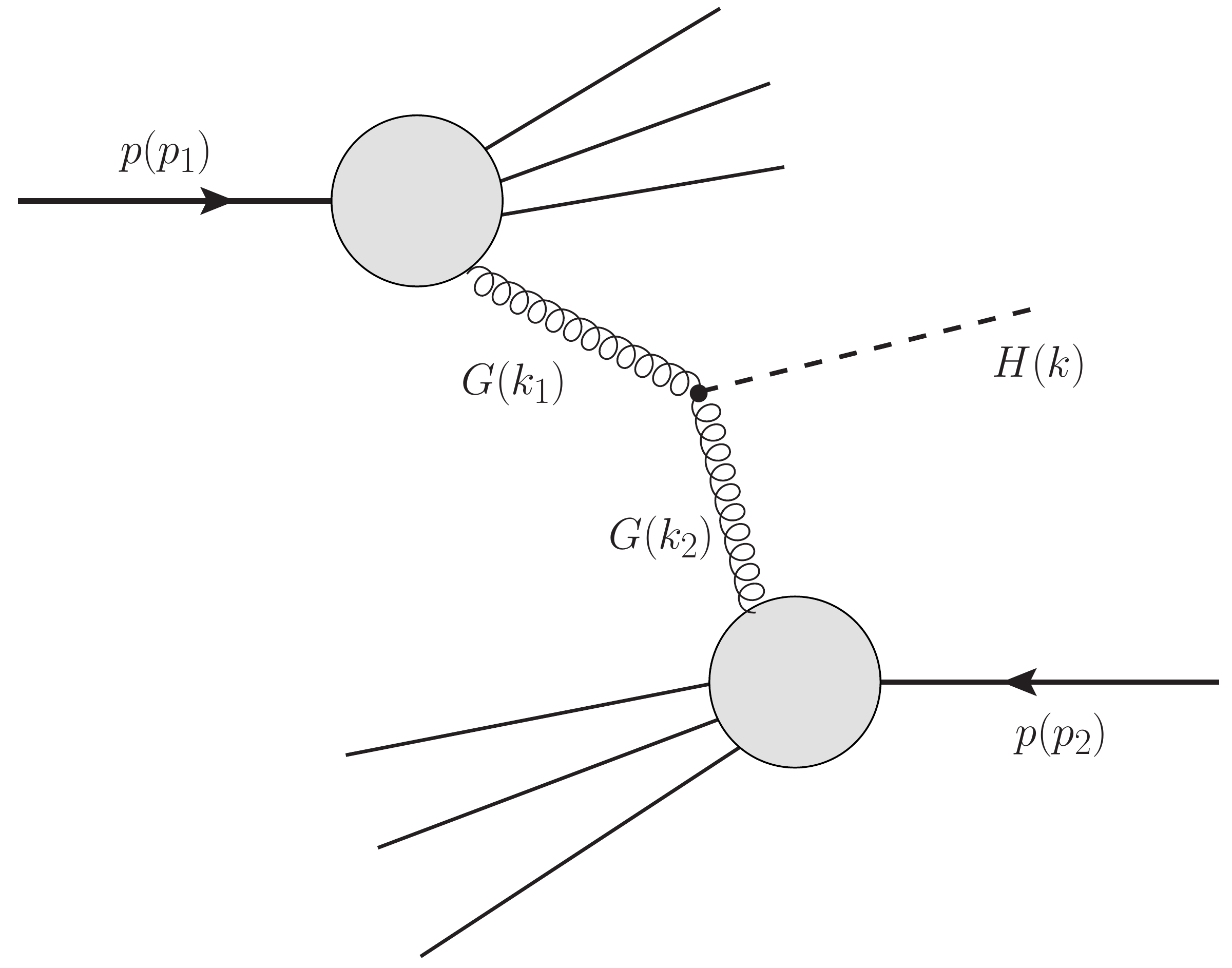}
\caption{Higgs-boson production in $pp$ collisions via gluon-gluon annihilation.}
\label{8Higgs-boson}
\end{center}
\end{figure}

Recently, at the LHC a boson was discovered \cite{200, 201} which, presumably, is the Higgs-boson 
$H_{SM}$ of the standard model of particle physics.
The main production mechanism for $H_{SM}$ in $pp$ collisions at LHC is gluon-gluon fusion; 
see figure \ref{8Higgs-boson}.
In this section we want to discuss the question if entanglement of the spins of the two gluons in 
figure \ref{8Higgs-boson} may influence the production rate of $H_{SM}$ or other scalar bosons.

We consider, thus, the reaction 
\be\label{4.1}
p(p_1)+p(p_2) \to H(k) + X
\ee
via gluon-gluon annihilation
\be\label{4.2}
G(k_1)+G(k_2) \to H(k), \qquad k_1+k_2 = k.
\ee
Here $H$ stands generically for a scalar boson.
We shall restrict ourselves to the collinear approximation and work in the $pp$ c.m.\ system.
We choose Cartesian unit vectors $\mbv{e}{1}, \mbv{e}{2}, \mbv{e}{3}$ with
\ba\label{4.3}
\mbv{e}{3} &=& \mbv{p}{1}/|\mbv{p}{1}|, \nn \\
\mbv{e}{i}\cdot \mbv{e}{j} &=& \delta_{ij}, \nn \\
(\mbv{e}{1}\times \mbv{e}{2})\cdot \mbv{e}{3} &=& 1.
\ea
We have then
\begin{alignat}{2}
\label{4.4}
 \mbv{p}{1} &= |\mbv{p}{1}|\mbv{e}{3}, &\qquad \mbv{p}{2} &=-|\mbv{p}{1}|\mbv{e}{3}, \nn \\
 \mbv{k}{1} &= x_1\mbv{p}{1}, &\qquad \mbv{k}{2} &= x_2\mbv{p}{2}.
\end{alignat}
We set 
\be\label{4.4a}
s = (p_1+p_2)^2 \
\ee
and assume high energies, that is,
\be\label{4.4b}
s \gg m^2_p. 
\ee
We have then, neglecting terms of relative order $m_p^2 / s$,
\begin{align}\label{4.4c}
 p_1^0 &= p_2^0 = \frac12 \sqrt{s}, \nn\\
 |\mbv{p}{1}| &= |\mbv{p}{2}| = \frac12 \sqrt{s}, \nn\\
 x_1 &= \frac{k^0 + k^3}{\sqrt{s}}, \nn \\
 x_2 &= \frac{k^0-k^3}{\sqrt{s}}, \nn \\
 m^2_H &= x_1 x_2 s. \
\end{align}

\subsection{Gluonic spin density matrices} 
Consider now an unpolarised proton $p(\mbv{p}{1})$ with a collinear gluon $G(\mbv{k}{1})$ in it.
Its spin and colour-spin density matrix must be of the form
\be\label{4.5}
\begin{split}
\rho^{(G)}_{a,i;a',i'} (x_1) &= \frac18 \delta_{aa'} \left[ a(x_1)\delta_{ii'} + b(x_1)\varepsilon_{ii'} \right], \\
 i,i' \in \{1,2\},&\quad a,a' \in \{1,\dots,8\}.  
\end{split}
\ee
Here we use rotational invariance around the axis $\mbv{e}{3}$, $a,a'$ are the colour-spin indices, $i,i'$ the spin indices in the linear polarisation basis, and
\be\label{4.6}
(\varepsilon_{ii'}) = \begin{pmatrix*}[r] 0&1 \\ -1& 0 \end{pmatrix*}.
\ee
Parity (P) invariance requires 
\be\label{4.6a}
b(x_1) = 0. 
\ee
Then, \eqref{4.5} together with the normalisation condition 
\be\label{4.6b}
\rho^{(G)}_{a,i;a',i'}(x_1)\delta_{aa'}\delta_{ii'} = 1 
\ee
implies
\be\label{4.6c}
\rho^{(G)}_{a,i;a',i'}(x_1) = \frac18 \delta_{aa'} \frac12 \delta_{ii'}.
\ee
Thus, the spin density matrix for a collinear gluon in an unpolarised proton is the trivial one.

Now we come to the general spin and colour-spin density matrix for the two gluons in \eqref{4.2}.
The two protons in \eqref{4.1} are supposed to be unpolarised and the gluons collinear to them.
We have rotational invariance around the $\mbv{e}{3}$ axis, P invariance, and we assume 
absence of colour correlations of the two gluons.
The most general density matrix reads then
\begin{align}\label{4.7}
  \rho^{(G,G)}_{a,b,i,j;a',b',i',j'}(x_1,x_2) 
  &= \frac{1}{64}\delta_{aa'}\delta_{bb'} \hat{\rho}^{(G,G)}_{i,j;i',j'}(x_1,x_2),
\\
\label{4.8a}
  & a,b,a',b' \in \{1,\dots ,8\}, \quad  i,j,i',j' \in \{1,2\} \nn
\end{align}
with 
\be
\label{4.8}
  \hat{\rho}^{(G,G)}_{i,j;i',j'}(x_1,x_2) =
   c_1(x_1,x_2)\delta_{ii'}\delta_{jj'} 
  + c_2(x_1,x_2)\delta_{ij} \delta_{i'j'} 
  +c_3(x_1,x_2)\delta_{ij'}\delta_{ji'}.
\ee                                                 
Here $a,i,a',i' \text{ and } b,j,b',j'$ refer to the colour-spin and spin indices of gluon 
$G(\mbv{k}{1})$ and $G(\mbv{k}{2})$, respectively.
The parameters of the density matrix are the real functions $c_l(x_1,x_2) \; (l=1,2,3)$.

For the further discussion it is useful to introduce the helicity basis.
We define for the gluon $G(\mbv{k}{1})$
\be\label{4.9}
 \begin{split}
\mbv{e}{+}^{(1)} &= -\frac{1}{\sqrt{2}}(\mbv{e}{1} + i\mbv{e}{2}), \\
\mbv{e}{-}^{(1)} &= \hphantom{-}\frac{1}{\sqrt{2}}(\mbv{e}{1} - i\mbv{e}{2})
 \end{split}
\ee
and for gluon $G(\mbv{k}{2})$
\be\label{4.10}
 \begin{split}
\mbv{e}{+}^{(2)} &= -\frac{1}{\sqrt{2}}(\mbv{e}{1} - i\mbv{e}{2}), \\
\mbv{e}{-}^{(2)} &= \hphantom{-}\frac{1}{\sqrt{2}}(\mbv{e}{1} + i\mbv{e}{2}).
 \end{split}
\ee
The density matrix $\hat{\rho}^{(G,G)}$, \eqref{4.8}, reads then
\begin{align}\label{4.11}
 \begin{split}
\hat\rho^{(G,G)}(x_1,x_2) &= \mbv{e}{i}\otimes\mbv{e}{j} \: \hat\rho^{(G,G)}_{i,j;i',j'}(x_1,x_2) \mbv{e}{i'}^\dagger \otimes \mbv{e}{j'}^\dagger \\
			  &= \mbv{e}{m}^{(1)}\otimes\mbv{e}{n}^{(2)} \hat\rho^{(G,G)}_{m,n;m',n'}(x_1,x_2) \mbv{e}{m'}^{(1)\dagger} \otimes \mbv{e}{n'}^{(2)\dagger},
 \end{split}
\end{align}
where
\be\label{4.12}
m,n,m',n' \in \{+,-\}.
\ee
From \eqref{4.8} to \eqref{4.11} we get $\rhog{m,n;m',n'}$ as follows
\begin{align}\label{4.13}
 \begin{split}
  \rhog{++;++}\xx &= \rhog{--;--}(x_1,x_2)\\
			&= c_1\xx + c_2\xx, \\
 \rhog{++;--}\xx &= \rhog{--;++}\xx \\
			&= c_2\xx + c_3\xx, \\
\rhog{+-;+-}\xx &= \rhog{-+;-+}\xx \\
			&= c_1\xx + c_3\xx, \\
 \end{split}
\end{align}
and all other matrix elements $\rhog{m,n;m',n'}\xx = 0$.
We still have the normalisation condition
\be\label{4.14}
\mathrm{Tr}\rhog{}\xx = 1
\ee
which implies
\be\label{4.15}
4c_1\xx + 2c_2\xx + 2c_3\xx = 1.
\ee
Thus, one of the functions $c_i\xx$ can be eliminated and we choose as independent 
parameters of $\rhog{}$
\begin{align}\label{4.16}
 \begin{split}
\xi\xx &= 2c_1\xx + 2c_2\xx, \\
\zeta\xx &= 2c_2\xx + 2c_3\xx.
 \end{split}
\end{align}
Together with \eqref{4.15} this gives
\begin{alignat}{2}\label{4.16a}
  c_1(x_1, x_2) = &+\frac14  \,&-\,\frac14\zeta\xx, \nn\\
  c_2\xx = &-\frac14 + \frac12\xi\xx \,&+\, \frac14\zeta\xx, \nn\\
  c_3\xx = &+\frac14 - \frac12\xi\xx \,&+\, \frac14\zeta\xx. 
\end{alignat}
With all this the matrix $\rhog{}$ reads in the helicity basis as shown in Table 3.

We still have the constraint that $\rhog{}$ must be a positive semi-definite matrix
\be\label{4.17}
\rhog{}\xx \geq 0.
\ee
This implies 
\be\label{4.18}
 \begin{split}
0 &\leq \xi\xx \leq 1, \\
-\xi\xx &\leq \zeta\xx \leq \xi\xx.
 \end{split}
\ee
Finally, for a reaction with identical parent hadrons, as is the case in \eqref{4.1}, we have
\be\label{4.19}
 \begin{split}
  \xi\xx &= \xi(x_2,x_1), \\
  \zeta\xx &= \zeta(x_2,x_1).
 \end{split}
\ee

\begin{table}[tb]\caption{The matrix elements $\rhog{m,n;m',n'}\xx$ parametrised by two real functions 
$\xi\xx$ and $\zeta\xx$.}
\centering
\begin{tabular}{r|c|c|c|c|}
$\begin{array}{c}
~~~~~m',n'\\
\raisebox{0pt}[0pt][0pt]{
{$\cdot$\raisebox{-0.3ex}
{$\cdot$}\raisebox{-0.7ex}
{$\cdot$}\raisebox{-1.1ex}
{$\cdot$}\raisebox{-1.5ex}
{$\cdot$}\raisebox{-1.9ex}
{$\cdot$}\raisebox{-2.3ex}
{$\cdot$}\raisebox{-2.7ex}
{$\cdot$}\raisebox{-3.1ex}
{$\cdot$}}}
\\ \\m,n~~~~~
\end{array}$
&$++$&$-~-$&$+-$&$-+$\\ \hline
&&&&\\
$++$~~~~~~&$\frac12\xi\xx$&$\frac12\zeta\xx$&$0$&$0$\\
&&&&\\ \hline 
&&&&\\
$-~-$~~~~~~&$\frac12\zeta\xx$&$\frac12\xi\xx$&$0$&$0$\\
&&&& \\ \hline 
&&&& \\
$+-$~~~~~~&$0$&$0$&$\frac12(1-\xi\xx)$&$0$\\
&&&& \\\hline
&&&&\\
$-+$~~~~~~&$0$&0&$0$&$\frac12(1-\xi\xx)$\\
&&&& \\ \hline
\end{tabular}

\label{Tab:3}
\end{table}

This is all we can say on general grounds about the density matrix $\rhog{}$. The trivial 
matrix corresponding to uncorrelated gluon spins is, of course, given by
\begin{align}\label{4.20}
 \begin{split}
 \left. \rhog{}\xx \right|_{\text{standard}} &= \frac14 \mathbbm{1}_4,\\ 
 \left. \xi\xx \right|_{\text{standard}} &= \frac12, \\
 \left. \zeta\xx \right|_{\text{standard}} &= 0.
 \end{split}
\end{align}
An example of a non-standard density matrix is discussed in appendix C.
We also note that here a \emph{factorising} two-gluon density matrix, with the 
one-gluon density matrices satisfying rotational and P invariance,
\emph{ must be of the standard form}, \eqref{4.20}, due to \eqref{4.6c}.

\subsection{Production of a $CP=+1$ Higgs boson in $pp$ collisions}

We consider now the reaction \eqref{4.1}, \eqref{4.2} for a $CP=+1$ Higgs boson $H$.
An example of $H$ is, of course, the SM Higgs boson.
From colour, Lorentz, CPT, and CP invariance we have
\be\label{4.21}
 \begin{split}
\langle H(k)|\cal T &|G(k_1,\varepsilon^{(1)},a),G(k_2,\varepsilon^{(2)},b)\rangle \\ 
 &= \braket{G(k_1,-\varepsilon^{(1)*},a),G(k_2,-\varepsilon^{(2)*},b)}{\cal T}{H(k)}\\
 &= -\delta_{ab} \frac{A}{m_H}\left[(\varepsilon^{(2)} \cdot \varepsilon^{(1)})(k_1 \cdot k_2)-(\varepsilon^{(2)} \cdot k_1)(\varepsilon^{(1)} \cdot k_2) \right],\\
 & a,b \in \left\{1,\dots,8\right\}.
 \end{split}
\ee
Here $k_{1,2} \text{ and } \varepsilon^{(1,2)}$ are the momenta and polarisation vectors of 
the gluons 1 and 2, respectively, and we have
\begin{align}\label{4.22}
 \begin{split}
  k_1 + k_2 &= k, \\
  k^2 &= m_H^2.
 \end{split}
\end{align}
Furthermore, $A$ is a dimensionless complex constant.

A standard calculation gives for the decay rate of $H$ into two gluons
\be\label{4.23}
\Gamma(H\to GG) = \frac1{8\pi}m_H |A|^2.
\ee
Another standard calculation gives for the transition rate of $GG \to H$ with the two gluons 
correlated according to the spin and colour spin density matrix \eqref{4.7}, \eqref{4.8}
\be\label{4.24}
\mathrm{d}\Gamma(G(k_1)+G(k_2)\to H(k))|_\rho 
= \frac1V \frac{1}{2k_1^0 2k_2^0} \frac{\mathrm{d}^3 k}{2k^0} 2\pi \delta^{(4)}(k-k_1-k_2)R,
\ee
\begin{align}\label{4.25}
 \begin{split}
  R = &\braket{H(k)}{\cal T}{G(k_1,\varepsilon_i^{(1)},a),G(k_2,\varepsilon_j^{(2)},b)} \\
      & \times \rho^{(G,G)}_{a,b,i,j;a',b',i',j'}\xx \\
      & \times \braket{H(k)}{\cal T}{G(k_1,\varepsilon_{i'}^{(1)},a'),G(k_2,\varepsilon_{j'}^{(2)},b')}^*.
 \end{split}
\end{align}
Here $V$ is the normalisation volume and we work in the $pp$ c.m.\ system, see \eqref{4.3}, \eqref{4.4}, 
where
\be\label{4.26}
\varepsilon^{(1)}_i = \begin{pmatrix} 0 \\ \mbv{e}{i} \end{pmatrix}, \qquad 
\varepsilon^{(2)}_j = \begin{pmatrix} 0 \\ \mbv{e}{j} \end{pmatrix}, \qquad
i,j \in \{1,2\}.
\ee
From \eqref{4.21} we have
\be\label{4.27}
\braket{H(k)}{\cal T}{G(k_1,\varepsilon_i^{(1)},a),G(k_2,\varepsilon_j^{(2)},b)} = \delta_{ab}\delta_{ij} \frac12 m_H A.
\ee
Inserting \eqref{4.7}, \eqref{4.8}, and \eqref{4.27} in \eqref{4.25} we get
\be\label{4.28}
R = \frac1{16} m_H^2 |A|^2 \left[c_1\xx+2c_2\xx+c_3\xx  \right].
\ee
With \eqref{4.16a} and \eqref{4.23} we obtain
\be\label{4.29}
R = \frac\pi4 m_H \Gamma (H\to GG) \left[\xi\xx + \zeta\xx \right].
\ee
Inserting this in \eqref{4.24} and using standard formulae from the parton model, 
see for instance chapter 18.5 of \cite{202}, we get the differential and total cross sections 
for reaction \eqref{4.1} as follows
\begin{align}\label{4.30}
 \begin{split}
  &\frac{\partial\sigma}{\partial k^3}(p(p_1)+p(p_2) \rightarrow H(k) + X) \\
  & = \frac{\pi^2}{4s} \frac{\Gamma(H\to GG)}{m_H k^0} N^p_G(x_1)N_G^p(x_2) \left[\xi\xx + \zeta\xx \right],
 \end{split}
\end{align}
\begin{align}\label{4.31}
 \begin{split}
  \sigma(p(p_1)+p(p_2)\to H + X) &= \frac{\pi^2}{4s}\frac{\Gamma(H\to GG)}{m_H} \\
				  & \times  \int^1_0\mathrm{d} x_1 \int^1_0\mathrm{d} x_2 \,\delta(x_1x_2 - \frac{m_H^2}{s}) \\
				   & \times  N_G^p (x_1) N_G^p (x_2) \left[\xi\xx + \zeta \xx \right].
 \end{split}
\end{align}

Note that in our collinear approximation the $\mbv{k}{T}$ distribution of the $H$ boson is proportional 
to $\delta^2(\mbv{k}{T})$ and has been integrated over in \eqref{4.30}.
Furthermore, $N_G^p (x)$ are the gluon distribution functions of the proton.
In \eqref{4.30} $x_1 \text { and } x_2 $ are to be inserted from \eqref{4.4c}.

The standard results at leading-order QCD are obtained from \eqref{4.30} and \eqref{4.31} by 
setting $\xi\xx = \frac12$ and $\zeta \xx = 0$; see \eqref{4.20}.
For the example of a correlated density matrix discussed in appendix C we have 
$\xi\xx = 1/2,\;\zeta\xx = 1/2$.
Inserting this in \eqref{4.30} and \eqref{4.31} we get \emph{twice} the standard results.
Note that the positivity constraints \eqref{4.18} allow from zero to four times the standard results.

\subsection{Production of a $CP=-1$ Higgs boson in $pp$ collisions}

In many models with an extended scalar sector there are both, $CP=+1$ scalar bosons $H$ and $CP=-1$ 
scalar bosons which we shall denote generically by $\tilde H$.
Of course, also scalar bosons with no definite $CP$ quantum number exist in various models.
The extension of our considerations to this case is straightforward.

We consider, thus, in this section the production of a $CP=-1$ scalar boson $\tilde H$ in $pp$ collisions 
via gluon-gluon fusion; see \eqref{4.1}, \eqref{4.2} with $H$ replaced by $\tilde H$.
From colour, Lorentz, CPT and CP invariance we find here (compare \eqref{4.21}):
\begin{align}\label{4.32}
 \begin{split}
  &\braket{\tilde H(k)}{\cal T}{G(k_1,\varepsilon^{(1)},a), G(k_2,\varepsilon^{(2)},b)} \\
  & = \braket{G(k_1,-\varepsilon^{(1)*},a),G(k_2,-\varepsilon^{(2)*},b)}{\cal T}{\tilde H (k)} \\
  & = -i\,\delta_{ab}\frac{\tilde A}{m_{\tilde H}}\varepsilon_{\mu\nu\rho\sigma}\varepsilon^{(1)\mu}\varepsilon^{(2)\nu}k_1^\rho k_2^\sigma,\\
  & \quad \,\, a,b \in \{1,\dots,8 \}.
 \end{split}
\end{align}
Here $\tilde A$ is a dimensionless complex constant and  
$\varepsilon_{\mu\nu\rho\sigma}\;(\varepsilon_{0123}=+1)$ is the totally antisymmetric tensor.
From this we find for the decay and production rates 
\be\label{4.33}
\Gamma(\tilde H \to GG ) = \frac{1}{8\pi} m_{\tilde H}|\tilde A|^2,
\ee
\be\label{4.34}
\mathrm{d}\Gamma \left.(G(k_1)+G(k_2) \to \tilde H(k)) \right|_\rho 
= \frac1V \frac{1}{2k^0_1 2k_2^0} \frac{\mathrm{d}^3 k}{2k^0} 2\pi \delta^{(4)}(k-k_1-k_2) \tilde R,\\
\ee
\begin{align}\label{4.35}
 \begin{split}
  \tilde R = &\braket{\tilde H(k)}{\cal T}{G(k_1,\varepsilon_i^{(1)},a),G(k_2,\varepsilon_j^{(2)},b)} \\
  & \; \times \rho^{(G,G)}_{a,b,i,j;a',b',i',j'}(x_1,x_2) \\
  & \times \braket{\tilde H (k)}{\cal T}{G(k_1,\varepsilon_{i'}^{(1)},a'),G(k_2,\varepsilon_{j'}^{(2)},b')}^*.  
 \end{split}
\end{align}
We find from \eqref{4.32} in the $pp$ c.m.\ system with \eqref{4.3}, \eqref{4.4}, \eqref{4.6}
\be\label{4.36}
\braket{\tilde H(k)}{\cal T}{G(k_1,\varepsilon_i^{(1)},a),G(k_2,\varepsilon_j^{(2)},b)} = i\delta_{ab}\frac12 m_{\tilde H} \tilde A \varepsilon_{ij},
\ee
\be\label{4.37}
 \begin{split}
  \tilde R &= \frac\pi4 m_{\tilde H} \Gamma (\tilde H \to GG) [2c_1\xx -2c_3\xx] \\
           & = \frac\pi4 m_{\tilde H} \Gamma (\tilde H \to GG) [\xi\xx - \zeta \xx]. 
 \end{split}
\ee
The differential and total cross sections read here as follows (compare \eqref{4.30}, \eqref{4.31}):
\be\label{4.38}
 \begin{split}
  &\frac{\partial\sigma}{\partial k^3}(p(p_1)+p(p_2) \rightarrow \tilde H(k) + X) \\
  & = \frac{\pi^2}{4s} \frac{\Gamma(\tilde H\to GG)}{m_{\tilde H} k^0} N^p_G(x_1)N_G^p(x_2) \left[\xi\xx - \zeta\xx \right],
 \end{split}
\ee
\be\label{4.39}
 \begin{split}
  \sigma(p(p_1)+p(p_2)\to \tilde H + X) =& \frac{\pi^2}{4s}\frac{\Gamma(\tilde H\to GG)}{m_{\tilde H}} \\
				  & \times \int^1_0\mathrm d x_1 \int^1_0\mathrm{d} x_2 \,\delta(x_1x_2 - \frac{m_{\tilde H}^2}{s}) \\
				   & \times   N_G^p (x_1) N_G^p (x_2) \left[\xi\xx - \zeta \xx \right].
 \end{split}
\ee
In \eqref{4.38} $x_1 \text{ and } x_2$ are to be inserted from \eqref{4.4c}.
Using \eqref{4.20} we obtain the standard leading-order QCD results from \eqref{4.38}, \eqref{4.39} 
for $\xi\xx = 1/2,\;\zeta\xx = 0$.
The example of the correlated density matrix of appendix C gives here zero cross sections.

To summarise: in this chapter we have shown that entanglement of the gluon spins can have a drastic 
influence on the production of $CP=+1 \text{ and } CP=-1 $ scalar bosons in $pp$ collisions via 
gluon-gluon annihilation.
This happens already for the collinear case.
A two-gluon density matrix factorising into single-gluon density matrices gives, 
in the collinear case, the standard results.
Thus, scalar-boson production via gluon-gluon annihilation is a sensitive probe of 
two-gluon entanglement effects.

\section{Conclusions}
\label{Conclusions}
In this paper we have first reviewed ideas on the QCD vacuum and how non-perturbative QCD 
effects may influence for instance the Drell-Yan process 
and soft photon production in hadron-hadron collisions.
In chapter 2 this was taken as motivation to discuss the Drell-Yan process with the ansatz 
of a general quark-antiquark density matrix, as suggested in \cite{1,2}.
We emphasise that this ansatz allows for both a factorising and a non-factorising, that is, 
entangled density matrix.
The TMD approach working with transverse momentum dependent parton distributions, see for 
instance \cite{5, 44b, 137, Boer:2002ju, 203, 204}, only allows a factorising density matrix of the 
form \eqref{3.28} to \eqref{3.30}.
We emphasise that our approach is more general than the TMD approach - which it includes as a special case - 
and was proposed earlier. In fact, our approach was developed in \cite{1} and \cite{2} {\em before} the authors 
had knowledge of the relevant experiments of \cite{6, 7}. To our knowledge it was in \cite{2} where 
for the first time it was shown 
that a non-trivial spin-transverse-momentum $q\bar q$ density matrix can have drastic effects for the 
lepton angular distributions in the unpolarised Drell-Yan reaction. Only later, in \cite{5}, the question 
of parton entanglement as an effect not describable in the TMD framework came to the forefront.
Our suggestion to experimentalists working on the Drell-Yan process is to use our general approach 
for the analysis.
In this way they may discover signs of \emph{parton entanglement} as was discussed in chapter 
\ref{Signatures of parton entanglement}.
Already a study of parton transverse momenta as extracted from the Drell-Yan reaction compared to, 
for instance, semi-inclusive-deep-inelastic scattering could be very interesting in this respect.
Even for enthusiasts of the TMD approach it should be very interesting to have a more general 
framework which allows to \emph{test} experimentally the basic assumption of a \emph{factorising} 
density matrix made there. The theoretical proofs of factorisation rely on QCD perturbation theory; see 
\cite{JC1,JC2} for reviews. These proofs clearly do not exclude violations of the factorisation 
hypothesis due to \emph{non-perturbative} QCD effects as discussed in \cite{122,1,2,3,4} and 
in the present article.

The question arises if and how the effects discussed in \cite{1, 2, 3, 4} and the present paper can be calculated in QCD. 
We can in this respect point to the calculations in the framework of the instanton approach; see \cite{122, 5, BRU}. 
A really satisfactory calculation certainly must simultaneously take into account 
the {\em hadron structure} and {\em in addition} the effects from the collisions; see points (i) and (v), 
respectively, of the summary of section 2. It is clear that this is a difficult task beyond the scope of the 
present paper. It may be that string-inspired methods, the AdS/CFT correspondence, could be applied to perform such calculations. 
For the original papers concerning this method see \cite{Maldacena:1997re, Gubser:1998bc, Witten:1998qj}, for a short 
introduction see \cite{Schomerus:2007ff}. The first applications of this method to problems of high energy scattering 
were done in \cite{Rho:1999jm, Janik:1999zk, Janik:2000aj}.
 
After the completion of the present paper J.-C. Peng kindly sent us the nice review \cite{Peng:2014hta}.
There, the emphasis is put on factorisation as basis for the analysis of the Drell-Yan reaction 
and related processes. In our paper, in contrast, we emphasise that factorisation may be - and in our 
opinion most probably is - violated due to non-perturbative QCD effects. 
Thus, these papers are rather complementary.

In chapter \ref{Higgs-boson production} we have discussed the production of 
$CP=+1 \text{ and } CP=-1$ scalar bosons $H \text{ and } \tilde H$, respectively, 
via gluon-gluon annihilation in $pp$ collisions.
We have shown that already in the collinear case gluon-gluon entanglement may drastically 
influence the differential and total cross sections for $H $ and $\tilde H$ production.
In this collinear case a factorising two-gluon density matrix must be the trivial one.
Thus, a careful study of Higgs-boson production at the LHC should allow to discover -- or 
at least to set limits on -- the gluon-gluon entanglement effects discussed here.

Going in our approach beyond the collinear case for gluon-gluon annihilation would be straightforward. We would make 
the ansatz of a general two-gluon density matrix containing all possible spin and transverse-momentum correlations. 
As a special case we would then find the TMD ansatz which was developed extensively in recent years following the 
initial paper \cite{Mulders:2000sh}; see \cite{Boer:2010zf, Qiu:2011ai, Sun:2011iw, Boer:2011kf, Boer:2013fca}. Clearly, 
a detailed investigation of the borderline between effects describable in the TMD framework and those requiring parton 
entanglement would be worthwhile. But this is beyond the scope of the present paper. Here we have restricted 
ourselves to the collinear case where the TMD approach gives no non-trivial effects but gluon entanglement 
may lead to {\em drastic} effects; see sections 4.2 and 4.3.

In this paper we have only been concerned with the Drell-Yan reaction and the Higgs-boson 
production for \emph{unpolarised} hadrons in the initial state.
The generalisation of the discussions to polarised hadrons would be straightforward.
For the Drell-Yan case the parameters $\mbv{F}{}$, $\mbv{G}{}$, $\underline{H}$ of the $q\bar q$ 
density matrix would then also depend on the spin parameters of the initial hadrons;
compare \eqref{3.7} and \eqref{A.20}, \eqref{A.20a}.
Clearly, also higher order QCD effects should be investigated.
But all this is beyond the scope of the present paper.

To summarise: we have pointed out that \cite{1} and \cite{2} were the first papers where non-trivial 
spin and transverse 
momentum density matrices for quarks and antiquarks in the Drell-Yan reaction were considered. 
In \cite{2} a detailed study of the influence of these non-trivial density matrices on the lepton 
angular distributions from production of a virtual photon and a $Z$ boson was given. 
Later, in \cite{5}, the question of parton entanglement versus the - in the meantime 
introduced - TMD factorisation came to the forefront. We think that a search for effects of parton 
entanglement in high-energy 
hadron-hadron collisions should be a very worth-while goal for experimentalists.
We have discussed the Drell-Yan reaction and Higgs-boson production in $pp$ collisions.
But, if entanglement effects exist, they should also show up in other reactions.
An example may be the production of quarkonium states in hadronic collisions.
We note that $0^{++} \text{ and } 2^{++}$ quarkonium states can be produced via gluon-gluon annihilation.
Thus, effects similar to those discussed for scalar bosons in section 4 may be important also there.

\section*{Acknowledgements}

The author would like to thank many colleagues for useful discussions and correspondence, in particular, 
I. Abt, A. Bacchetta, P. Bordalo, O. Denisov, M. Diehl, C. Ewerz, O. Eyser, W. Hollik, N. Makins, 
S. Paul, J.-C. Peng, M. Radici, P.E. Reimer, O. Teryaev, and P. Zavada.
Special thanks are due to the organisers of the ECT* workshop ``Drell-Yan Scattering and the 
Structure of Hadrons'' from 21 to 25 May 2012, for creating there such a nice and stimulating atmosphere.
The author has profited a lot from the talks and discussions at this workshop. Special thanks are also due 
to E. Bittner, C. Ewerz, and S. Casas for help in preparing the manuscript.

\begin{appendix}

\section{The Drell-Yan reaction with general $q\bar q$ density matrix}

Here we give the detailed formulae for the Drell-Yan reaction
\begin{align}\label{A.1}
\begin{split}
h_1(p_1) + h_2(p_2) \to &V(k) + X \\
			& \drsh l^{+}(q_1) + l^{-}(q_2)
\end{split}
\end{align}
with $V=\gamma^* \text{ or } Z$ for a general density matrix \eqref{3.7} of the annihilating $q\bar q$ pair.
We use the definitions and notations of \cite{2} except for the replacements 
$\mbv{e}{i}^* \to \mbv{e}{i}'$ $(i=1,2,3)$,
$\mbv{p}{i}^* \to \mbv{p}{i}'$ $(i=1,2)$. etc.

We note first that the general density matrix $\rho^{(q,\bar q)}$ \eqref{3.7} must be positive semi definite
\be\label{A.100}
\rho^{(q,\bar q)} \geq 0.
\ee
For the elements of the upper-left $2\times 2 $ part of $\rho^{(q,\bar q)}$ which is relevant 
for the Drell-Yan reaction (see table 2) this implies in particular
\be\label{A.101}
 \begin{split}
 &1\geq H_{33} \geq -1, \\
 &(1+H_{33})^2-(F_3+G_3)^2 - (H_{11}-H_{22})^2 - (H_{12}+H_{21})^2 \geq 0.
\end{split}
\ee

In \eqref{3.25} to \eqref{3.33} of \cite{2} the $V$ production matrix at parton level 
is defined and its general expansion is given:
\be\label{A.0}
 \begin{split}
r^V_{ij}(\mbv{k}{1}';\rho;q\bar q) =&\, r^{V,ij}(\mbv{k}{1}';\rho;q\bar q) \\
 =& \,\braket{V(i)}{\cal T}{q(\mbv{k}{1}',\alpha;A),\bar q(-\mbv{k}{1}',\beta;B) } \\
  & \,\times \rho^{(q,\bar q)}_{\alpha\beta,\alpha'\beta'} \frac19 \delta_{AA'} \delta_{BB'} \\
 & \,\times \braket{V(j)}{\cal T}{q(\mbv{k}{1}',\alpha';A'),\bar q(-\mbv{k}{1}',\beta';B') }^* \\
 =& \,c^V \left\{ \frac13 \delta_{ij} \tilde a^V + \frac{1}{2i} \varepsilon_{ijl} \tilde B^{V,l} - \tilde C ^{V,ij} \right\}.
 \end{split}
\ee
Here $A,A',B,B' \in \{1,2,3\}$ are the colour indices of $q \text{ and } \bar q$.
The Cartesian polarisation indices, $i,j \in \{1,2,3\}$, of the vector boson $V$ refer to the coordinate 
axes \eqref{3.6} as defined in the $q\bar q$ c.m.\ system which is also
the rest system of $V$ in the leading order calculations considered here.
The coefficients $\tilde a^V,\, \tilde B^{V,l} \text{ and } \tilde C^{V,ij}$ occurring in \eqref{A.0} 
are given for $V=Z$ in (3.27) to (3.30) of \cite{2} for an arbitrary factorising $q\bar q$ density 
matrix (3.22) of \cite{2}.
Note that there is a misprint in (3.30) of \cite{2}. The correct equation for $\tilde C^Z$ reads
\begin{align}\label{A.01}
 \begin{split}
  \tilde C^{Z,ij}(\mbv{e}{3}',\mbv{s}{},\mbv{r}{},q)\equiv& \,\tilde C^{Z}_{ij}(\mbv{e}{3}',\mbv{s}{},\mbv{r}{},q) \\
  =& \, (e'^i_{3}e'^j_{3} - \frac13 \delta_{ij} )\left\{ \frac12 (g^2_{Vq}+g^2_{Aq})[1+(\mbv{s\cdot e}{3}')(\mbv{r}{}\cdot \mbv{e}{3}')] \right.\\
  & \quad\, - g_{Vq}g_{Aq} \mbv{e}{3}' \cdot (\mbv{s+r}{})+\frac12 (g^2_{Vq}-g^2_{Aq})  \\
  & \quad\, \times \left.[\mbv{s\cdot r}{}-(\mbv{s\cdot e_3}{}')(\mbv{r\cdot e_3}{}')]\phantom{\frac12}\!\!\!\right\} \\
  & \quad\, + [(s^{i}-e'^i_{3} (\mbv{s\cdot e}{3}'))(r^{j}-e'^j_{3} (\mbv{r\cdot e}{3}')) \\
  & \quad\, +(s^{j}-e'^j_{3} (\mbv{s\cdot e}{3}'))(r^{i}-e'^i_{3} (\mbv{r\cdot e}{3}')) \\
  & \quad\, -\frac23 \delta_{ij} (\mbv{s\cdot r}{}-(\mbv{s\cdot e}{3}')(\mbv{r\cdot e}{3}'))] \frac12 (g^2_{Vq}-g^2_{Aq}).
 \end{split}
\end{align}

From (3.25) to (3.29) of \cite{2} and \eqref{A.01} we get the $Z$ production matrix at parton 
level for the $q\bar q $ density matrix \eqref{3.7}
by  making the following replacements.
In the linear terms in $\mbv{s}{}$ and $\mbv{r}{}$ we have
\be\label{A.2}
 \begin{split}
  \mbv{s}{} \to \mbv{F}{} , \\
  \mbv{r}{} \to \mbv{G}{}.
 \end{split}
\ee
In the bilinear terms we have
 \be\label{A.3}
 \mbv{s \otimes r}{} \to H_{ij} \mbv{e}{i}' \otimes \mbv{e}{j}'.
 \ee
 In this way we obtain for quark-flavour $q$
 \be\label{A.4}
 c^Z = \frac{1}{12} \frac{e^2 m_Z^2}{\sin^2 \theta_W \cos^2 \theta_W},
 \ee
 \begin{align}\label{A.5}
    \tilde a^Z_q &= (g_{Vq}^2 + g_{Aq}^2)(1+H_{33}) - 2g_{Vq}g_{Aq} (F_3+G_3),\\
  \mbv{\tilde B}{q}^Z &= -\mbv{e}{3}' [2g_{Vq}g_{Aq}(1+H_{33}) - (g_{Vq}^2 + g_{Aq}^2)(F_3+G_3) ],\label{A.6}
\end{align}
\be\label{A.7}
 \begin{split}
  \tilde C^{Z,ij}_q = & \,({e'}_3^{i} {e'}_3^{j} - \frac13 \delta_{ij}) \bigg[ \frac12 (g_{Vq}^2 + g_{Aq}^2) (1+H_{33}) - g_{Vq}g_{Aq} (F_3+G_3) \bigg] \\
		      & \! +  \frac12 (g_{Vq}^2 - g_{Aq}^2) \left[ ({e'}_1^{i} {e'}_1^{j} - {e'}_2^{i} {e'}_2^{j}) (H_{11}-H_{22}) + ({e'}_1^{i} {e'}_2^{j} + {e'}_2^{i} {e'}_1^{j})(H_{12}+H_{21}) \right].
 \end{split}
\ee
Here we have
\begin{align}\label{A.7a}
\begin{split}
&g_{Vq} = T_{3q} - 2Q_q \sin^2 \theta_W , \\              
&g_{Aq} = T_{3q},\\ 
\end{split}
\intertext{with $\theta_W$ the weak mixing angle and } 
\begin{split}
& T_{3q} = 1/2, \; Q_q = 2/3 \quad \text{ for $u$-type quarks,} \\ 
& T_{3q} = -1/2, \; Q_q = -1/3 \quad \text{ for $d$-type quarks.} 
\end{split}\label{A.7b}
\end{align}

Comparing with table \ref{Tab:2}  we find that for the massless quark flavours $q$ considered here only the matrix elements of $\rho^{(q, \bar q)}$ with $RL \text{ and } LR$ enter in \eqref{A.5} to \eqref{A.7}
as it must be due to \eqref{3.20}.
We have
\begin{align}\label{A.8}
 \begin{split}
  1+H_{33} &= 2(\rho_{RL,RL}^{(q, \bar q)}+ \rho_{LR,LR}^{(q, \bar q)}),\\
  F_3+G_3 &= 2(\rho_{RL,RL}^{(q, \bar q)} - \rho_{LR,LR}^{(q, \bar q)}), \\
  H_{11}-H_{22} &= 2(\rho_{RL,LR}^{(q, \bar q)}+ \rho_{LR,RL}^{(q, \bar q)}), \\
  H_{12}+H_{21} &= 2i (\rho_{RL,LR}^{(q, \bar q)} - \rho_{LR,RL}^{(q, \bar q)}).  
 \end{split}
\end{align}

For the ordinary Drell-Yan process, $V=\gamma^*$, we have to make the following replacements 
in \eqref{A.4} to \eqref{A.7}:
\be\label{A.9}
 \begin{split}
  m_Z &\to m_{\gamma^*} ,\\
  \frac{e}{\sin\theta_W \cos\theta_W} &\to e, \\
  g_{Vq} &\to 2Q_q, \\
  g_{Aq} &\to 0.
 \end{split}
\ee
This gives
\begin{align}\label{A.10}
 \begin{split}
 c^{\gamma^*} &= \frac{1}{12} e^2 m_{\gamma^*}^2, \\
 \tilde a^{\gamma^*}_q &= 4Q_q^2(1+H_{33}),\\
 \mbv{\tilde B}{q}^{\gamma^*} &= 4Q^2_q \mbv{e}{3}'(F_3+G_3), \\
  \tilde C^{\gamma^*,ij}_q & = 2Q_q^2 \left[ ({e'}_3^{i} {e'}_3^{j} - \frac13 \delta_{ij}) (1+H_{33}) \right.
 \\
 &\hspace*{1.5cm}
 +  ({e'}_1^{i} {e'}_1^{j} - {e'}_2^{i} {e'}_2^{j}) (H_{11}-H_{22}) 
 \\
 &\hspace*{1.5cm} \left.
+ ({e'}_1^{i} {e'}_2^{j} + {e'}_2^{i} {e'}_1^{j})(H_{12}+H_{21}) \right].
 \end{split}
\end{align}

It is easy to write the partonic production matrix \eqref{A.0} and  \eqref{A.4} to \eqref{A.7} in a covariant way.
We consider here massless quarks where $k_1^2 =k_2^2 =0$.
We set
\be\label{A.10a}
 \begin{split}
  K &= k_1+k_2, \\
  \hat s &= K \cdot K = 2k_1 \cdot k_2, \\ 
  s &= (p_1+p_2)^2 \\
 \end{split}
\ee
and we have then
\be
  k_1 \cdot K  = k_2 \cdot K = \frac12 \hat s.
\label{A.11}
\ee
We define now 
\be
 \begin{split}
  \varepsilon_0^\mu & = K^\mu \frac{1}{\sqrt{\hat s}} \;, \\
  \varepsilon^\mu_3 & = (k_1^\mu - k_2^\mu) \frac{1}{\sqrt{\hat s}} \;, 
 \end{split}\label{A.12}
\ee
\be\label{A.13}
\begin{split}
 \varepsilon_1^\mu & = \varepsilon^{\mu\nu\rho\sigma} (p_1+p_2)_\nu k_{1\rho}k_{2\sigma} \xi , \\
 \xi &= \left[\hat s ((p_1+p_2) \cdot k_1) ((p_1+p_2) \cdot k_2) - \frac{s\hat s^2}{4} \right]^{-1/2}, 
\end{split}
\ee
\be\label{A.14}
\varepsilon_2^\mu = \varepsilon^{\mu\nu\rho\sigma} \varepsilon_{3\nu} \varepsilon_{1\rho} K_{\sigma}\frac{1}{\sqrt{\hat s}}
\ee
where we use the normalisation $\varepsilon_{0123} = +1$ for the totally antisymmetric 
$\varepsilon_{\mu\nu\rho\sigma}$ symbol.
In the c.m.\ system of the $q\bar q$ collision we have
\begin{align}
\label{A.15} \begin{split}
  (\varepsilon^\mu_0) &= \begin{pmatrix} 1 \\ 0 \end{pmatrix} , \\
  (\varepsilon^\mu_a) &= \begin{pmatrix} 0 \\ \mbv{e}{a}' \end{pmatrix} , \quad \text{ for $a=1,2,3$; }
 \end{split}
\end{align}
see \eqref{3.6}.

The partonic production matrix reads now in covariant form, using \eqref{A.0} and \eqref{A.4} 
to \eqref{A.7}, as follows
\begin{align}\label{A.16}
\begin{split}
r^{V,\mu\nu} (k_1,k_2;\rho;q\bar q) =& \,\braket{V^\mu (k)}{\cal T}{q(k_1,\alpha,A), \bar q (k_2,\beta, B)} \\
				    & \times \rho^{(q, \bar q)}_{\alpha \beta, \alpha'\beta'}\frac19 \delta_{AA'}\delta_{BB'} \\
				    & \times \braket{V^\nu(k)}{\cal T}{q(k_1,\alpha',A'), \bar q (k_2,\beta', B')}^* , \\
				     V =& \gamma^*, Z.
\end{split}
 \end{align}
For $Z$ production we have
 \begin{align}\label{A.17}
\begin{split}
r^{Z,\mu\nu} (k_1, k_2; \rho; q\bar q) = \frac12 c^Z \bigg\{& \Big[ (g_{Vq}^2+g_{Aq}^2)(1+H_{33}) - 2g_{Vq}g_{Aq} (F_3+G_3) \Big] \Big[ -g^{\mu\nu} + \varepsilon_0^\mu \varepsilon_0^\nu - \varepsilon_3^\mu \varepsilon_3^\nu\Big ]\\
							    + & \Big[- 2g_{Vq}g_{Aq} (1+H_{33}) +(g_{Vq}^2+g_{Aq}^2) (F_3+G_3)  \Big] \frac{2i}{\hat s} \varepsilon^{\mu\nu\rho\sigma} k_{1\rho}k_{2\sigma} \\
							    - & (g_{Vq}^2-g_{Aq}^2) \Big[(H_{11}-H_{22})(\varepsilon_1^\mu \varepsilon_1^\nu - \varepsilon_2^\mu \varepsilon_2^\nu) \\
							    + &  (H_{12}+H_{21})(\varepsilon_1^\mu \varepsilon_2^\nu + \varepsilon_2^\mu \varepsilon_1^\nu) \Big] \bigg\}.
\end{split}
 \end{align}
The partonic production matrix for $\gamma^*$ in covariant form is obtained from \eqref{A.17} by 
making the replacements \eqref{A.9}:
\begin{align}\label{A.17b}
\begin{split}
r^{\gamma^*,\mu\nu} (k_1, k_2; \rho; q\bar q) = \frac{2\pi}{3}\alpha m_{\gamma^*}^2 Q_q^2 \bigg\{& (1+H_{33})(-g^{\mu\nu} + \varepsilon_0^\mu \varepsilon_0^\nu - \varepsilon_3^\mu \varepsilon_3^\nu) \\
							    + & (F_3+G_3)  \frac{2i}{\hat s} \varepsilon^{\mu\nu\rho\sigma} k_{1\rho}k_{2\sigma} \\
							    - & \Big[(H_{11}-H_{22})(\varepsilon_1^\mu \varepsilon_1^\nu - \varepsilon_2^\mu \varepsilon_2^\nu) \\
							    + &  (H_{12}+H_{21})(\varepsilon_1^\mu \varepsilon_2^\nu + \varepsilon_2^\mu \varepsilon_1^\nu) \Big] \bigg\}.
\end{split}
 \end{align}
In the $q \bar q$ c.m.\ system we have
\begin{equation}\label{A.17a}
\left(r^{V,\mu\nu}(k_1,k_2;\rho;q\bar q)\right)=
\left(
\begin{array}{c|c}
  0 & 0 \\ \hline \\[-9pt]
  0 & r^{V,ij} (\mbv{k}{1}';\rho;q\bar q)
\end{array}
\right)
\end{equation}
with $r^{V,ij} (\mbv{k}{1}';\rho;q\bar q)$ from \eqref{A.0}.

Here it is appropriate to discuss again the functional dependences of the ``unconventional'' 
parameters $F_3, G_3, H_{ij}$ occurring in \eqref{A.17} and \eqref{A.17b}.
These parameters are discussed here for massless quarks and, in general, will depend on the quark flavour $q$.
We have given the parity properties of $\mbv{F}{}$, $\mbv{G}{}$ and  $\underline{H}$ in \eqref{3.11}.
From this we find that $F_3$ and $G_3$ must be P-odd, that is, proportional to the only P-odd invariant we 
can form from the four vectors $p_1, p_2, k_1 \text{ and } k_2$ available:
\be\label{A.18}
I = \varepsilon_{\mu\nu\rho\sigma}p_1^\mu p_2^\nu k_1^\rho k_2^\sigma.
\ee
Note that \emph{all four vectors} are needed to form $I$.
The P-even parameters available are (see \eqref{3.1} and \eqref{3.4})
\be\label{A.19}
J = \left\{ s, x_1, x_2, \mbv{k}{1T}^2, \mbv{k}{2T}^2, \mbv{k}{1T}\cdot \mbv{k}{2T}\right\}.
\ee
Thus, in general, we get
\begin{align}\label{A.20}
 \begin{split}
  F_3 &= If_3^{(q)} (J)\;,\\
  G_3 &= Ig_3^{(q)} (J) \;,\\
  \end{split}
  \end{align}
  and
  \begin{align}\label{A.20a}
    \begin{split}
  H_{33} &= H_{33}^{(q)} (J) \;, \\
  H_{11}-H_{22} &= (H_{11}-H_{22})^{(q)} (J) \;, \\
  H_{12}+H_{21} &= I (h_{12} + h_{21})^{(q)} (J).
  \end{split}
  \end{align}
Note that we have factored out the P-odd invariant $I$ in $(H_{12}+H_{21})$ since $\varepsilon_1$ is P-odd, 
$\varepsilon_2$ is P-even, but the tensor 
\be\label{A.21}
(H_{12}+H_{21})(\varepsilon_1^\mu \varepsilon_2^\nu + \varepsilon_2^\mu \varepsilon_1^\nu)
\ee
occurring in \eqref{A.17}, \eqref{A.17b} must be P-even; see also \eqref{3.9}, \eqref{3.11}.
In the collinear case,
\be\label{A.22}
\mbv{k}{1T} = \mbv{k}{2T} = 0,
\ee
we have rotational symmetry around the $q\bar q$ collision axis in the $q \bar q$ c.m.\ system.
This implies, together with P invariance,
\begin{align}\label{A.23}
 \begin{split}
  F_3 = G_3 &= 0 \;,\\
  H_{11}-H_{22} & = 0 \;,\\
  H_{12}+H_{21} &= 0 \;,
 \end{split}
\end{align}
for $\mbv{k}{1T}=\mbv{k}{2T}=0$.
Note that only $H_{33}$ survives in this case which corresponds to the situation already discussed 
in \cite{1}.

From $r^{V,\mu\nu}$ \eqref{A.16} we get the overall production matrix of the $V$ boson in 
the $h_1$-$h_2$ collision by integration over the quark and antiquark momentum distributions in the hadrons.
As in \cite{2} we allow for $\mbv{k}{T}$ correlations of $q \text{ and } \bar q$.
We define the $V$ production matrix as follows
\begin{align}\label{A.24}
 \begin{split}
  R^{V,\mu\nu} (h_1, h_2; k) =& \sum_X (2\pi)^4 \delta^{(4)} (p_1+p_2-k-k_X) \\
			      & \times \sideset{}{'}\sum_{h_1,h_2 \text { spins}} \!\!\!\!\!\braket{V^\mu(k), X(k_X)}{\cal T}{h_1(p_1), h_2(p_2)} \braket{V^\nu(k), X(k_X)}{\cal T}{h_1(p_1), h_2(p_2)}^*
 \end{split}
\end{align}
where $\sum'$ means the average over the spins of $h_1$ and $h_2$.
See also (2.3) of \cite{2}.
We get then
\begin{align}\label{A.25}
 \begin{split}
  R^{V,\mu\nu}(h_1, h_2; k) = \sum_q \bigg\{& \int\limits_0^1 \mathrm{d}x_1\, \frac{N_q^{h_1}(x_1)}{x_1} \int\limits_0^1\mathrm{d}x_2\,\frac{N_{\bar q}^{h_2}(x_2)}{x_2} \\
			     & \times \int\mathrm{d}^2 k_{1T}\,\mathrm{d}^2 k_{2T} \; P_{12} (\mbv{k}{1T},\mbv{k}{2T}) (2\pi)^4 \delta^{(4)} (k-k_1-k_2) r^{V,\mu\nu}(k_1,k_2;\rho;q\bar q) \\
			     & + \text{an analogous term with $\bar q$ from $h_1$ and $q$ from $h_2$ } \bigg\}.
\end{split}
\end{align}
Here $k_{1,2}$ are the quark and antiquark momenta as given in the overall c.m.\ system in \eqref{3.4} 
and $N_q^{h_1}(x_1)$, $N_{\bar q}^{h_2}(x_2)$ are the standard parton distribution functions.
In \eqref{A.25} $P_{12} (\mbv{k}{1T},\mbv{k}{2T})$ is the $q\bar q$ transverse-momentum distribution. 
In \cite{2} a simple ansatz was made, allowing for transverse momentum correlations,
\begin{align}\label{A.26}
 \begin{split}
   P_{12} (\mbv{k}{1T},\mbv{k}{2T}) &= \frac{\alpha_T(\alpha_T+2\beta_T)}{\pi^2} 
				   \exp \left[-\alpha_T(\mbv{k}{1T}^2+\mbv{k}{2T}^2)-\beta_T(\mbv{k}{1T}-\mbv{k}{2T})^2 \right],\\
				   & \alpha_T >0, \quad \beta_T >-\frac12\alpha_T.
 \end{split}
\end{align}
Here $\alpha_T$ and $\beta_T$ are parameters which could depend on $s, x_1 \text{ and } x_2$.
For $\beta_T\neq 0$ the transverse momenta of $q$ and $\bar q$ are correlated.
From the ansatz \eqref{A.26} we get the mean squares of the $q, \bar q$  and the vector boson's 
transverse momenta, $\mbv{k}{1T}, \mbv{k}{2T}$ and $\mbv{k}{T} = \mbv{k}{1T}+\mbv{k}{2T}$, 
respectively, as follows:
\begin{align}\label{A.27}
 \begin{split}
  \langle \mbv{k}{T}^2 \rangle &= \frac{2}{\alpha_T}\;,\\
  \langle \mbv{k}{1T}^2 \rangle &= \langle \mbv{k}{2T}^2 \rangle = \frac{\alpha_T+\beta_T}{\alpha_T(\alpha_T+2\beta_T)}\\
				&= \frac12 \langle \mbv{k}{T}^2 \rangle \frac{1+\beta_T/\alpha_T}{1+2\beta_T/\alpha_T}.  
 \end{split}
\end{align}
For no correlation, $\beta_T=0$, this gives
\be\label{A.28}
\langle \mbv{k}{1T}^2 \rangle = \langle \mbv{k}{2T}^2 \rangle = \frac12 \langle \mbv{k}{T}^2 \rangle.
\ee
For maximal positive correlation, $\beta_T/\alpha_T\to \infty$, we get, however,
\be\label{A.29}
\langle \mbv{k}{1T}^2 \rangle = \langle \mbv{k}{2T}^2 \rangle = \frac14 \langle \mbv{k}{T}^2 \rangle.
\ee
Let us suppose now that in nature the transverse momenta of $q$ and $\bar q$ are indeed highly correlated.
Then, an estimate of $\langle \mbv{k}{1T}^2 \rangle$ and $\langle \mbv{k}{2T}^2 \rangle$ 
from the observed $\langle \mbv{k}{T}^2 \rangle$ of the vector boson and assuming no correlation,
that is using \eqref{A.28}, will give a value of the partonic transverse momenta which is too 
large by a factor up to 2; see \eqref{A.27} and \eqref{A.29}.

In the ansatz \eqref{A.26} we have chosen a function symmetric under the 
exchange $\mbv{k}{1T}\leftrightarrow \mbv{k}{2T}$.
Clearly, this could easily be made more general allowing for different mean squared transverse momenta of 
quarks and antiquarks in $h_1$ and $h_2$.

To write down the cross section for the whole reaction \eqref{A.1} of $V$ production with subsequent 
leptonic decay we still need the decay matrices for $V\to l^+ l^-$.
These are defined as
\be\label{A.30}
  D^V_{\mu\nu} (q_1,q_2) = \sum_{\alpha, \beta}  \braket{l^+(q_1,\alpha), l^-(q_2, \beta)}{\cal T}{V_{\mu}(k)}^* 
						\braket{l^+(q_1,\alpha), l^-(q_2, \beta)}{\cal T}{V_{\nu}(k)}
\ee
where $\alpha \text{ and } \beta$ are the spin indices of $l^+$ and $l^-$, respectively, and we 
assume no observation of the lepton polarisations.
These matrices are given in the $V$ rest system in appendix A of \cite{2}.
From (A.5) of \cite{2} we find, in covariant notation, for $V=Z$ setting $m_l = 0$:
\begin{align}\label{A.31}
 \begin{split}
  D^{Z}_{\mu\nu} (q_1,q_2) =& \,48\pi \, m_Z\, \Gamma(Z\to l^+l^-) \\
			    & \times \Big\{ -\frac12 g_{\mu\nu} + \frac{1}{m_Z^2}(q_{1\mu}q_{2\nu}+q_{2\mu}q_{1\nu}) 
			    + \frac{i}{m_Z^2}\varepsilon_{\mu\nu\rho\sigma}\, q_1^\rho q_2^\sigma \frac{2 g_{Vl}g_{Al}}{g_{Vl}^2+g_{Al}^2}\Big\}.
 \end{split}
\end{align}
Here 
\begin{align}\label{A.32}
 \begin{split}
  g_{Vl} &= -\frac12 + 2\sin^2 \theta_W \;, \\
  g_{Al} &= -\frac12 \;,\\
  \Gamma(Z\to l^+ l^-) &= \frac{\alpha m_Z (g_{Vl}^2+g_{Al}^2)}{12 \sin^2\theta_W \cos^2\theta_W}.
 \end{split}
\end{align}
The cross section for the reaction \eqref{A.1} with $V=Z$ is then
\begin{align}\label{A.33}
 \begin{split}
  \mathrm{d}\sigma (h_1+h_2 \to Z+X \to l^+ + l^- + X) =& \frac{1}{2w (s,m_1^2, m_2^2)} \frac{\mathrm{d^3} q_1 \mathrm{d^3} q_2}{(2\pi)^6 2q_1^0 2q_2^0}\\
					      &\times \left[(k^2-m^2_Z)^2 + m_Z^2\Gamma_Z^2 \right]^{-1}
					        D^Z_{\nu\mu}(q_1,q_2) R^{Z,\mu\nu} (h_1, h_2;k),\\
					      k=q_1+q_2.&
 \end{split}
\end{align}
Here $R^{Z,\mu\nu}$ is defined in \eqref{A.24}, \eqref{A.25}, $m_{1,2}$ are the masses of $h_{1,2}$, and 
\be\label{A.34}
w(x,y,z) = \left(x^2+y^2+z^2 -2xy-2yz-2xz\right)^{1/2}.
\ee
We have described the $Z$ line shape by a simple Breit-Wigner formula.
For $V=\gamma^*$ in \eqref{A.1} we get for the decay matrix from (A.8) of \cite{2} and \eqref{A.31}
\be\label{A.35}
D^{\gamma^*}_{\mu\nu} (q_1,q_2) = 8\pi\alpha \left[ -(q_1+q_2)^2 g_{\mu\nu} + 2(q_{1\mu}q_{2\nu}+q_{2\mu}q_{1\nu}) \right].
\ee
The cross section reads
\begin{align}\label{A.36}
 \begin{split}
  \mathrm{d}\sigma (h_1+h_2 \to \gamma^*+X \to l^+ + l^- + X) &= \frac{1}{2w (s,m_1^2, m_2^2)} \frac{\mathrm{d^3} q_1 \mathrm{d^3} q_2}{(2\pi)^6 2q_1^0 2q_2^0}\\
					      & \hphantom{=\;\;} \times \left(\frac{1}{k^2}\right)^2 D^{\gamma^*}_{\nu\mu}(q_1,q_2) R^{\gamma^*,\mu\nu} (h_1, h_2;k),\\
					      k=q_1+q_2,& \qquad
					      k^2 = m_{\gamma^*}^2.
 \end{split}
\end{align}
We note that from \eqref{A.35} we have
\be\label{A.37}
D^{\gamma^*}_{\mu\nu}(q_1, q_2) = D^{\gamma^*}_{\nu\mu}(q_1, q_2).
\ee
Therefore, in the contraction with $R^{\gamma^*\!,\mu\nu}$ in \eqref{A.36} the 
antisymmetric part of the latter drops out.
Looking at \eqref{A.25} and \eqref{A.17b} we see that from the correlation effects $(F_3 + G_3)$ will 
drop out and the Drell-Yan reaction with $V = \gamma^*$ and without observation of the lepton polarisations 
is only sensitive to
$(1+H_{33})$, $(H_{11}-H_{22})$, and $(H_{12}+H_{21})$.

Finally we remark that the lepton angular distributions following from \eqref{A.33} and \eqref{A.36} are 
guaranteed to satisfy all general positivity constraints of \cite{OT} since our $q\bar q$ density matrix
\eqref{3.7} is required to be positive semi definite; see \eqref{A.100}, \eqref{A.101}.

If the lepton polarisations could be observed one would get access also to $(F_3+G_3)$ in the ordinary 
Drell-Yan process.
Indeed, suppose that we could select exclusively leptons $l^+$ with longitudinal polarisation 
$r_1 /2$ where $r_1 \in \{-1,1\}$.
Then the corresponding decay matrix of the $\gamma^*$ reads, neglecting the lepton mass,
\be\label{A.38}
  D^{\gamma^*}_{\mu\nu} (q_1, r_1; q_2) = 4\pi\alpha \left\{ -g_{\mu\nu}(q_1+q_2)^2 
							    {} + 2(q_{1\mu}q_{2\nu} + q_{2\mu}q_{1\nu})
							   + 2i r_1 \varepsilon_{\mu\nu\rho\sigma} q_1^\rho q_2^\sigma \right\}.
\ee
This has an antisymmetric part giving a non-zero result when contracted with the antisymmetric 
part in $r^{\gamma^*\!, \mu\nu}$ which is proportional to $(F_3+G_3)$; see \eqref{A.17b}.

With this we close our review of the kinematic formulae for the reaction \eqref{A.1}.
These formulae are in essence from \cite{2} but are written here in covariant form.
We note that there also is a $\gamma^*$-$Z$ interference term which could be easily written down 
for our general $q \bar q$ density matrix with the methods presented here. 
 
\section{The angular distribution of the lepton pair in the New Trento Convention.}
 
 In \cite{21} a new convention for the notation of momenta and the definition of angular variables is given.
We discuss here how this influences the angular distribution \eqref{3.17}.
In the New Trento Convention (TR) the angles $\theta_{TR}$ and $\phi_{TR}$ refer to the $l^-$ 
momentum in the Collins-Soper frame, 
whereas we used in \cite{2} and in \eqref{3.17} the angles $\theta$ and $\phi$ of the $l^+$ momentum.
Thus, we have
\begin{align}\label{B.1}
 \begin{split}
  \theta &= \pi - \theta_{TR}\;, \\
  \phi & = \pi + \phi_{TR}. 
 \end{split}
\end{align}
This gives
\begin{align}\label{B.2}
 \begin{split}
  \sin \theta  &= \sin \theta_{TR}\;,\\
  \cos\theta &= -\cos\theta_{TR}\;,\\
  \sin(2\theta) &= -\sin(2\theta_{TR})\;,\\
  \cos\phi &= -\cos \phi_{TR} \;,\\
  \cos(2\phi) & = \cos (2\phi_{TR})\;.
 \end{split}
\end{align}
Inserting this in \eqref{3.17} we see that the angular distribution looks exactly the same in our 
convention from \cite{2} and in the New Trento Convention.

\section{An example of a non-standard two-gluon spin density matrix}

Let us assume that the two gluons annihilating to give the boson $H$ in figure \ref{8Higgs-boson} 
have correlated transverse polarisation.
As an example we first consider completely correlated transverse polarisation of the two gluons 
with the three-dimensional polarisation vectors of both gluons given by
\be\label{C.1}
\mbv{e}{}(\phi) = \cos \phi \; \mbv{e}{1} + \sin\phi\; \mbv{e}{2}.
\ee
Here we use the coordinate system \eqref{4.3}, \eqref{4.4}.
The spin part of the two-gluon density matrix \eqref{4.7}, \eqref{4.8}, \eqref{4.11} is 
then constructed by integrating over $\phi$
\be\label{C.2}
\rhog{}\xx = \frac{1}{2\pi} \int\limits_0^{2\pi} \mathrm{d}\phi \;\; \mbv{e}{}(\phi)\otimes \mbv{e}{}(\phi) \;\; \mbv{e}{}^\dagger(\phi) \otimes \mbv{e}{}^\dagger(\phi).
\ee
The integral in \eqref{C.2} is easily performed and gives a density matrix as shown in table \ref{Tab:3} with
\begin{align}\label{C.3}
 \begin{split}
  \xi\xx  &= \frac12 \;,\\
  \zeta\xx &= \frac12 \;.
 \end{split}
\end{align}

We emphasise that this exercise is only meant to give an \emph{example} how a non-trivial two-gluon density matrix could be built up.
Background fields as discussed in section \ref{The QCD vaccum structure}, for instance instantons, may do such a job.
But this remains to be investigated in detail.
The density matrix with the parameters as in \eqref{C.3} is certainly \emph{not} to be cited as ``the prediction of our model''.

\end{appendix}


\begin{thebibliography}{99}
\bibitem{1}
  O.~Nachtmann and A.~Reiter,
  ``The Vacuum Structure in QCD and Hadron - Hadron Scattering,''
  Z.\ Phys.\ C {\bf 24} (1984) 283.

\bibitem{2}
  A.~Brandenburg, O.~Nachtmann and E.~Mirkes,
  ``Spin effects and factorization in the Drell-Yan process,''
  Z.\ Phys.\ C {\bf 60} (1993) 697.

\bibitem{3}
  G.~W.~Botz, P.~Haberl and O.~Nachtmann,
  ``Soft photons in hadron hadron collisions: Synchrotron radiation from the QCD vacuum?,''
  Z.\ Phys.\ C {\bf 67} (1995) 143
  [hep-ph/9410392].

\bibitem{4}
  O.~Nachtmann,
  ``The QCD vacuum structure and its manifestations,''
  in `` On Confinement Physics,'' eds. S.~D.~Bass and P.~A.~M. Guichon, Editions Frontieres, 1996.

\bibitem{5}
  D.~Boer, A.~Brandenburg, O.~Nachtmann and A.~Utermann,
  ``Factorisation, parton entanglement and the Drell-Yan process,''
  Eur.\ Phys.\ J.\ C {\bf 40} (2005) 55
  [hep-ph/0411068].

\bibitem{6}
  S.~Falciano {\it et al.}  [NA10 Collaboration],
  ``Angular distributions of muon pairs produced by 194-GeV/c negative pions,''
  Z.\ Phys.\ C {\bf 31} (1986) 513.

\bibitem{7}
  M.~Guanziroli {\it et al.}  [NA10 Collaboration],
  ``Angular distributions of muon pairs produced by negative pions on deuterium and tungsten,''
  Z.\ Phys.\ C {\bf 37} (1988) 545.

\bibitem{8} J.~S.~Conway {\em et al.}, ``Experimental study of muon pairs produced by 252-GeV pions 
on tungsten,'' Phys.\ Rev.\ D {\bf 39} (1989) 92.

\bibitem{9}
  J.~G.~Heinrich, {\it et al.},
  ``Higher-twist effects in the reaction $\pi^- N\to \mu^+\mu^- X$ at 253-GeV/c,''
  Phys.\ Rev.\ D {\bf 44} (1991) 1909.

\bibitem{10}
 W.~Erni {\it et al.} [PANDA Collaboration], 
``Physics Performance Report for PANDA: Strong Interaction Studies with Antiprotons,'' 
arXiv: 0903.3905 [hep-ex].

\bibitem{11}
  V.~Barone {\it et al.}  [PAX Collaboration],
  ``Antiproton-proton scattering experiments with polarization,''
  arXiv: hep-ex/0505054.

\bibitem{12}
M.~Diefenthaler, The E906/SeaQuest experiment at Fermilab, Projects Document 1265-v1, 
PANIC11, Cambridge, Massachusetts, July 2011.

\bibitem{13}
COMPASS-II Proposal, CERN-SPSC-2010-014; 
SPSC-P-340, \\{\em http://cdsweb.cern.ch/record/1265628/files/SPSC-P-340.pdf}

\bibitem{14}
L.~Y.~Zhu, {\em et al.}, ``Measurement of Angular Distributions of Drell-Yan Dimuons 
in $p + d$ Interactions at 800 GeV/c,'' Phys. Rev. Lett. {\bf 99}, (2007) 082301.  

\bibitem{15}
L.~Y.~Zhu, {\em et al.}, ``Measurement of Angular Distributions of Drell-Yan Dimuons in $p + p$ 
Interactions at 800 GeV/c,'' Phys. Rev. Lett. {\bf 102}, (2009) 182001. 

\bibitem{16}
O.~Eyser, ``Drell Yan @ RHIC'', Talk at the ECT* Workshop on Drell Yan Physics and the Structure 
of Hadrons, May 21-25, 2012, Trento, Italy.

\bibitem{17}
T.~Aaltonen {\em et al.}, ``First Measurement of the Angular Coefficients of Drell-Yan $e^+e^-$ 
Pairs in the $Z$ Mass Region from 
$p\bar p$ Collisions at $\sqrt{s}=1.96$ TeV,'' 
The CDF Collaboration, Phys. Rev. Lett. {\bf 106}, (2011) 241801, arXiv: 1103.5699.

\bibitem{18}
F. Carminati, (Ed.) {\it et al.}  [ALICE Collaboration], ``ALICE: Physics performance report, volume I,''
  J.\ Phys.\ G {\bf 30} (2004) 1517.

\bibitem{19}
[ATLAS Collaboration], ``ATLAS: Letter of intent for a general purpose p p experiment 
at the large hadron collider at CERN,''
  CERN-LHCC-92-04 (1992).

\bibitem{20}
M.~Della Negra {\it et al.}  [CMS Collaboration],
``CMS: The Compact Muon Solenoid: Letter of intent for a general purpose detector 
at the LHC,'' Techn. Rep. CERN-LHCC-92-03, CERN-LHCC-1-1, Cern 1992.

\bibitem{21}
N.~C.~R.~Makins, ``Summary of Experimental Drell-Yan Physics,'' Talk at the 
ECT* Workshop on Drell-Yan Scattering and the Structure of Hadrons, May 21-25, 2012, Trento, Italy.

\bibitem{110}
  A.~A.~Belavin, A.~M.~Polyakov, A.~S.~Schwartz and Y.~.S.~Tyupkin,
  ``Pseudoparticle Solutions of the Yang-Mills Equations,''
  Phys.\ Lett.\ B {\bf 59} (1975) 85.

\bibitem{111}
  G.~'t Hooft,
 ``Symmetry Breaking Through Bell-Jackiw Anomalies,''
  Phys.\ Rev.\ Lett.\  {\bf 37} (1976) 8.

\bibitem{112}
  G.~'t Hooft,
  ``Computation of the Quantum Effects Due to a Four-Dimensional Pseudoparticle,''
  Phys.\ Rev.\ D {\bf 14} (1976) 3432
   [Erratum-ibid.\ D {\bf 18} (1978) 2199].

\bibitem{113}
  G.~K.~Savvidy,
  ``Infrared Instability of the Vacuum State of Gauge Theories and Asymptotic Freedom,''
  Phys.\ Lett.\ B {\bf 71} (1977) 133.

\bibitem{114}
  A.~I.~Vainshtein, V.~I.~Zakharov and M.~A.~Shifman,
  ``Gluon Condensate And Lepton Decays Of Vector Mesons.''
  JETP Lett.\  {\bf 27} (1978) 55
   [Pisma Zh.\ Eksp.\ Teor.\ Fiz.\  {\bf 27} (1978) 60].

\bibitem{115}
M.~A.~Shifman, A.~I.~Vainshtein and V.~I.~Zakharov,
  ``QCD and Resonance Physics. Theoretical Foundations,''
  Nucl.\ Phys.\ B {\bf 147} (1979)  385.

\bibitem{115a}
M.~A.~Shifman, A.~I.~Vainshtein and V.~I.~Zakharov,
  ``QCD and Resonance Physics. Applications,''
  Nucl.\ Phys.\ B {\bf 147} (1979) 448.  
    
\bibitem{116a}
  J.~Ambj\o rn and P.~Olesen,
  ``On the Formation of a Random Color Magnetic Quantum Liquid in QCD,''
  Nucl.\ Phys.\ B {\bf 170} (1980) 60.
  
\bibitem{116}
  J.~Ambj\o rn and P.~Olesen,
  ``A Color Magnetic Vortex Condensate in QCD,''
  Nucl.\ Phys.\ B {\bf 170} (1980) 265.

\bibitem{117}
J.~C.~Maxwell: Philos. Mag. {\bf 21}, 281 (1861), reproduced in: The Scientific 
Papers of James Clerk Maxwell, Vol. 1, p. 488, W.~D.~Niven, ed., Cambridge Univ. Press 1890. 

\bibitem{118}
  H.~M.~Fried and B.~Muller, (Eds.),
  ``QCD vacuum structure,'' Proceedings, Workshop, Paris, France, June 1-5, 1992, 
  Singapore: World Scientific (1993). 

\bibitem{119}
  A.~Donnachie, H.~G.~Dosch, P.~V.~Landshoff, and O.~Nachtmann,
``Pomeron physics and QCD,''
  Camb.\ Monogr.\ Part.\ Phys.\ Nucl.\ Phys.\ Cosmol.\  {\bf 19} (2002) 1.

\bibitem{120}
  S.~D.~Drell and T.~-M.~Yan,
  ``Massive Lepton Pair Production in Hadron-Hadron Collisions at High-Energies,''
  Phys.\ Rev.\ Lett.\  {\bf 25} (1970) 316
   [Erratum-ibid.\  {\bf 25} (1970) 902].

\bibitem{121}
  S.~D.~Drell and T.~-M.~Yan,
  ``Partons and their Applications at High-Energies,''
  Annals Phys.\  {\bf 66} (1971) 578
   [Annals Phys.\  {\bf 281} (2000) 450].

\bibitem{121a}
R.~K.~Ellis, W.~J.~Stirling, and B.~Webber,
``QCD and Collider Physics,''
Cambridge \ Univ. \ Press, Cambridge, \ U.K. (1996).

\bibitem{122}
J.~Ellis, M.~K.~Gaillard, W.~J.~Zakrzewski, 
``Nonperturbative effects on factorization in high-momentum processes,'' 
Phys.\ Lett.\  {\bf B81} (1979) 224.

\bibitem{123}
  H.~G.~Dosch,
  ``Nonperturbative methods in quantum chromodynamics,''
  Prog.\ Part.\ Nucl.\ Phys.\  {\bf 33} (1994) 121.

\bibitem{124}
 S.~Narison,
  ``QCD spectral sum rules,''
  World Sci.\ Lect.\ Notes Phys.\  {\bf 26} (1989) 1.

\bibitem{125}
  A.~Di Giacomo, H.~G.~Dosch, V.~I.~Shevchenko and Yu.~A.~Simonov,
  ``Field correlators in QCD: Theory and applications,''
  Phys.\ Rept.\  {\bf 372} (2002) 319
  [hep-ph/0007223].

\bibitem{126}
I.~M.~Ternov, Yu.~M.~Loskutov, L.~I.~Korovina, ``On the Possibility 
of Polarization of an Electron Beam due to Relativistic Radiation 
in a Magnetic Field,'' Zh. Eskp, Teor. Fiz. {\bf 41} (1961) 1294.

\bibitem{127}
 A.~A.~Sokolov and I.~M.~Ternov,
  ``On polarization and spin effects in the theory of synchrotron radiation,''
  Sov.\ Phys.\ Dokl.\  {\bf 8} (1964) 1203
   [Dokl.\ Akad.\ Nauk Ser.\ Fiz.\  {\bf 153} (1963) 1052].
   
\bibitem{128}
 J.~D.~Jackson,
  ``On Understanding Spin-Flip Synchrotron Radiation and the Transverse Polarization of Electrons in Storage Rings,''
  Rev.\ Mod.\ Phys.\  {\bf 48} (1976) 417.

\bibitem{129}
M.~D'Eglia, A.~Di Giacomo and E.~Meggiolaro,
  ``Field strength correlators in full QCD,''
  Phys.\ Lett.\ B {\bf 408} (1997) 315
  [hep-lat/9705032].

\bibitem{42a}
  D.~A.~Smith {\it et al.}  [UKQCD Collaboration],
  ``Topological structure of the SU(3) vacuum,''
  Phys.\ Rev.\ D {\bf 58} (1998) 014505
  [hep-lat/9801008].
  
\bibitem{42b}
  F.~Schrempp and A.~Utermann,
  ``QCD instantons and high-energy diffractive scattering,''
  Phys.\ Lett.\ B {\bf 543} (2002) 197
  [hep-ph/0207300].

\bibitem{130}
 J.~Antos {\it et al.},
  ``Soft photon production in 400-GeV/c~ $p$ - Be collisions,''
  Z.\ Phys.\ C {\bf 59} (1993) 547.

\bibitem{131}
D.~de Florian {\it et al.}, 
``QCD Spin Physics: Partonic Spin Structure of the Nucleon,''
  Prog.\ Part.\ Nucl.\ Phys.\  {\bf 67} (2012) 251
  [arXiv:1112.0904 [hep-ph]].

\bibitem{44b}
  D.~Boer, M.~Diehl  {\it et al.},
 ``Gluons and the quark sea at high energies: Distributions, polarization, tomography,'' 
 arXiv:1108.1713 [nucl-th].

\bibitem{44c}
M.~Diehl, ``Imaging partons in exclusive scattering processes,'' in I.~C.~Brock (Ed.), 
Proc. XXth Int. Workshop on Deep-Inelastic Scattering and Related Subjects,
 DIS 2012, 26-30 March 2012, Bonn, Germany,
 DESY-PROC-2012-02,
  [arXiv:1206.0844 [hep-ph]].
  
\bibitem{133}
  J.~Cortes and B.~Pire,
  ``About the vacuum structure in QCD and hadron - hadron scattering,''
  Z.\ Phys.\ C {\bf 29} (1985) 51.

\bibitem{134}
  C.~S.~Lam and W.~-K.~Tung,
  ``A Parton Model Relation Sans QCD Modifications In Lepton Pair Productions,''
  Phys.\ Rev.\ D {\bf 21} (1980) 2712.

\bibitem{135}
  E.~Mirkes and J.~Ohnemus,
  ``Angular distributions of Drell-Yan lepton pairs at the Tevatron: Order $\alpha_{s}^{2}$ corrections and Monte Carlo studies,''
  Phys.\ Rev.\ D {\bf 51} (1995) 4891
  [hep-ph/9412289].

\bibitem{136}
  P.~Chiappetta and M.~Le Bellac,
  ``Angular Distribution Of Lepton Pairs In Drell-Yan Like Processes,''
  Z.\ Phys.\ C {\bf 32} (1986) 521.

\bibitem{137}
D. Boer, ``Investigating the origins of transverse spin asymmetries at RHIC,'' 
Phys. Rev. D {\bf 60}, (1999) 014012.

\bibitem{BRU}
A. Brandenburg, A. Ringwald, A. Utermann, ``Instantons in lepton pair production,''  
Nucl. Phys. B {\bf 754}, (2006) 107.
  
\bibitem{200}
ATLAS Coll., ``Observation of a new particle in the search for the Standard Model Higgs boson with the 
ATLAS detector at the LHC,'' Phys. Lett. B {\bf 716}, (2012) 1 .

\bibitem{201}
CMS Coll., ``Observation of a new boson at a mass of 125 GeV with the CMS experiment 
at the LHC,'' Phys. Lett. B {\bf 716}, (2012) 30.

\bibitem{202}
O. Nachtmann, 
``Elementary Particle Physics, Concepts and Phenomena,'' 
Springer \ Verlag, Berlin, Heidelberg, 1990.

\bibitem{Boer:2002ju}
  D.~Boer, S.~J.~Brodsky and D.~S.~Hwang,
  ``Initial state interactions in the unpolarized Drell-Yan process,''
  Phys.\ Rev.\ D {\bf 67} (2003) 054003
  [hep-ph/0211110].
  
\bibitem{203}
P.~Mulders, ``Theory overview of TMDs,'' in I.~C.~Brock (Ed.), 
Proc. XXth Int. Workshop on Deep-Inelastic Scattering and Related Subjects,
 DIS 2012, 26-30 March 2012, Bonn, Germany,
 DESY-PROC-2012-02.

\bibitem{204}
M.~Radici, ``Theoretical Summary,'' Talk at the ECT* Workshop on Drell-Yan 
Scattering and the Structure of Hadrons, May 21 to 25, 2012, Trento, Italy.

\bibitem{JC1} 
J. C. Collins, ``Foundations of Perturbative QCD,'' Cambridge University Press, Cambridge, 2011.

\bibitem{JC2} 
J. C. Collins, ``TMD theory, factorization and evolution,'' 
Int. J. Mod. Phys. Conf. Ser. {\bf 25} (2014) 1460001 [arXiv:1307.2920].

\bibitem{Maldacena:1997re}
  J.~M.~Maldacena,
  ``The Large N limit of superconformal field theories and supergravity,''
  Adv.\ Theor.\ Math.\ Phys.\  {\bf 2} (1998) 231 [hep-th/9711200].
  
\bibitem{Gubser:1998bc}
  S.~S.~Gubser, I.~R.~Klebanov and A.~M.~Polyakov,
  ``Gauge theory correlators from noncritical string theory,''
  Phys.\ Lett.\ B {\bf 428} (1998) 105
  [hep-th/9802109].
  
\bibitem{Witten:1998qj}
  E.~Witten,
  ``Anti-de Sitter space and holography,''
  Adv.\ Theor.\ Math.\ Phys.\  {\bf 2} (1998) 253
  [hep-th/9802150].
  
\bibitem{Schomerus:2007ff}
  V.~Schomerus,
  ``Strings for Quantumchromodynamics,''
  Int.\ J.\ Mod.\ Phys.\ A {\bf 22} (2007) 5561
   [Conf.\ Proc.\ C {\bf 060726} (2006) 151]
  [arXiv:0706.1209 [hep-ph]].
  
\bibitem{Rho:1999jm}
  M.~Rho, S.~-J.~Sin and I.~Zahed,
  ``Elastic parton-parton scattering from AdS / CFT,''
  Phys.\ Lett.\ B {\bf 466} (1999) 199
  [hep-th/9907126].
  
\bibitem{Janik:1999zk}
  R.~A.~Janik and R.~B.~Peschanski,
  ``High-energy scattering and the AdS / CFT correspondence,''
  Nucl.\ Phys.\ B {\bf 565} (2000) 193
  [hep-th/9907177].
  
\bibitem{Janik:2000aj}
  R.~A.~Janik and R.~B.~Peschanski,
  ``Minimal surfaces and Reggeization in the AdS / CFT correspondence,''
  Nucl.\ Phys.\ B {\bf 586} (2000) 163
  [hep-th/0003059].

\bibitem{Peng:2014hta}
  J.~-C.~Peng and J.~-W.~Qiu,
  ``Novel Phenomenology of Parton Distributions from the Drell-Yan Process,''
  Prog.\ Part.\ Nucl.\ Phys.\  {\bf 76} (2014) 43
  [arXiv:1401.0934 [hep-ph]].
  
\bibitem{Mulders:2000sh}
  P.~J.~Mulders and J.~Rodrigues,
  ``Transverse momentum dependence in gluon distribution and fragmentation functions,''
  Phys.\ Rev.\ D {\bf 63} (2001) 094021
  [hep-ph/0009343].
  
\bibitem{Boer:2010zf}
  D.~Boer, S.~J.~Brodsky, P.~J.~Mulders and C.~Pisano,
  ``Direct Probes of Linearly Polarized Gluons inside Unpolarized Hadrons,''
  Phys.\ Rev.\ Lett.\  {\bf 106} (2011) 132001
  [arXiv:1011.4225 [hep-ph]].
  
\bibitem{Qiu:2011ai}
  J.~-W.~Qiu, M.~Schlegel and W.~Vogelsang,
  ``Probing Gluonic Spin-Orbit Correlations in Photon Pair Production,''
  Phys.\ Rev.\ Lett.\  {\bf 107} (2011) 062001
  [arXiv:1103.3861 [hep-ph]].
  
\bibitem{Sun:2011iw}
  P.~Sun, B.~-W.~Xiao and F.~Yuan,
  ``Gluon Distribution Functions and Higgs Boson Production at Moderate Transverse Momentum,''
  Phys.\ Rev.\ D {\bf 84} (2011) 094005
  [arXiv:1109.1354 [hep-ph]].
  
\bibitem{Boer:2011kf}
  D.~Boer, W.~J.~den Dunnen, C.~Pisano, M.~Schlegel and W.~Vogelsang,
  ``Linearly Polarized Gluons and the Higgs Transverse Momentum Distribution,''
  Phys.\ Rev.\ Lett.\  {\bf 108} (2012) 032002
  [arXiv:1109.1444 [hep-ph]].
  
\bibitem{Boer:2013fca}
  D.~Boer, W.~J.~den Dunnen, C.~Pisano and M.~Schlegel,
  ``Determining the Higgs spin and parity in the diphoton decay channel,''
  Phys.\ Rev.\ Lett.\  {\bf 111} (2013) 3,  032002
  [arXiv:1304.2654 [hep-ph]].

\bibitem{OT}
O.~Teryaev, ``Positivity constraints for quarkonia polarization,`` Nucl. Phys. B (Proc. Suppl.) {\bf 214} (2011) 118.

\end{thebibliography}
\end{document}